\documentclass[a4paper,fleqn]{cas-sc}

\usepackage[authoryear,longnamesfirst]{natbib}
\usepackage{color}
\usepackage{enumitem}
\def\tsc#1{\csdef{#1}{\textsc{\lowercase{#1}}\xspace}}
\tsc{WGM}
\tsc{QE}
\tsc{EP}
\tsc{PMS}
\tsc{BEC}
\tsc{DE}

\usepackage{amssymb,amsthm}
\theoremstyle{definition}
\newtheorem{proposition}{Proposition}
\newtheorem{lemma}{Lemma}

\newtheorem{rmk}{Remark}

\usepackage{booktabs}

\usepackage{bm}

\newcommand{\E}{{\mathbb E}}

\newcommand{\bA}{{\bm A}}
\newcommand{\cA}{{\cal A}}
\newcommand{\cG}{{\cal G}}
\newcommand{\cN}{{\cal N}}

\newcommand{\cV}{{\cal V}}
\newcommand{\cE}{{\cal E}}
\newcommand{{\cF}}{{\cal F}}
\newcommand{\cL}{{\cal L}}
\newcommand{\cX}{{\cal X}}
\newcommand{\cY}{{\cal Y}}
\newcommand{\cZ}{{\cal Z}}

\newcommand{\var}{ {\rm{var}} }

\usepackage{pdflscape}
\usepackage{float}
\usepackage{array}

\graphicspath{{figures/}}

\begin{document}
\let\WriteBookmarks\relax
\def\floatpagepagefraction{1}
\def\textpagefraction{.001}
\shorttitle{Supply Network Structure and the Bullwhip Effect}
\shortauthors{J. Y\"{u}, C. Cai, and J. Gao}

\title [mode = title]{Bullwhip Effect of Supply Networks: Joint Impact of Network Structure and Market Demand}                      



\author[1,2]{Jin-Zhu Y\"{u}}
\author[3]{Chencheng Cai}

\author[1,2]{Jianxi Gao}[orcid=0000-0002-3952-208X]

\cormark[1]
\ead{jianxi.gao@gmail.com}
\cortext[cor1]{Corresponding author}

\address[1]{Department of Computer Science, Rensselaer Polytechnic Institute (RPI), Troy, NY 12180, USA}
\address[2]{Network Science and Technology Center, Rensselaer Polytechnic Institute (RPI), Troy, NY 12180, USA}
\address[3]{Department of Statistical Science, Temple University, Philadelphia, PA 19122, USA}

\begin{abstract}
The progressive amplification of fluctuations in demand as the demand travels upstream the supply chains is known as the bullwhip effect. We first analytically characterize the bullwhip effect in general supply chain networks in two cases: (i) all suppliers have a unique layer position, where our method is founded on the control-theoretic approach, and (ii) not all suppliers have a unique layer position due to the presence of intra-layer links or inter-layer links between suppliers that are not positioned in consecutive layers, where we use both the absorbing Markov chain and the control-theoretic approach. We then investigate how network structures impact the BWE of supply chain networks. In particular, we analytically show that (i) if the market demand is generated from the same stationary process, the structure of supply networks does not affect the layer-wise bullwhip effect of supply networks, and (ii) if the market demand is generated from different stationary or non-stationary market processes, wider supply networks lead to a lower level of layer-wise bullwhip effect. Finally, numerical simulations are used to validate our propositions. 
\end{abstract}




\maketitle

\section{Introduction} \label{sec:intro}

Supply chain networks (henceforth referred to as supply networks) are essential to national and global economies by providing a sustained flow of products~\citep{gross2018introduction}. However, as a large-scale complex dynamical system, supply networks are prone to instabilities caused by disturbances \citep{gao2016universal,liu2022network,liu2016breakdown,li2021ripple}, such as market demand fluctuations in market demand. Due to the crucial role that supply networks play in the regional and global economies, it is critical to understand their response when they are subject to volatility of market demand.
The response of supply networks can be dynamic and stochastic, perhaps most notably manifested by the well-known bullwhip effect (BWE), i.e., the progressive amplification of fluctuations in flow (demand) as it travels upstream the supply chain~\citep{lee1997bullwhip}. BWE has a significant impact on supply chains: it causes dynamic inventory instabilities (stockouts or stockovers) and thus lead to excessive inventory and supply chain cost, which is reported in multiple industries~\citep{shan2014empirical,wolter2018quantifying,osadchiy2021bullwhip}. 
To develop effective mitigation strategies, it is critical to understand how the market demand volatility, together with the network structure, lead to the BWE of different supply networks. We seek to answer the following questions: How to characterize the BWE of general supply networks? How does the BWE of supply networks changes as the structure changes under different types of market demand?

\begin{figure}[htb]
\centering
\includegraphics[width=0.75\textwidth]{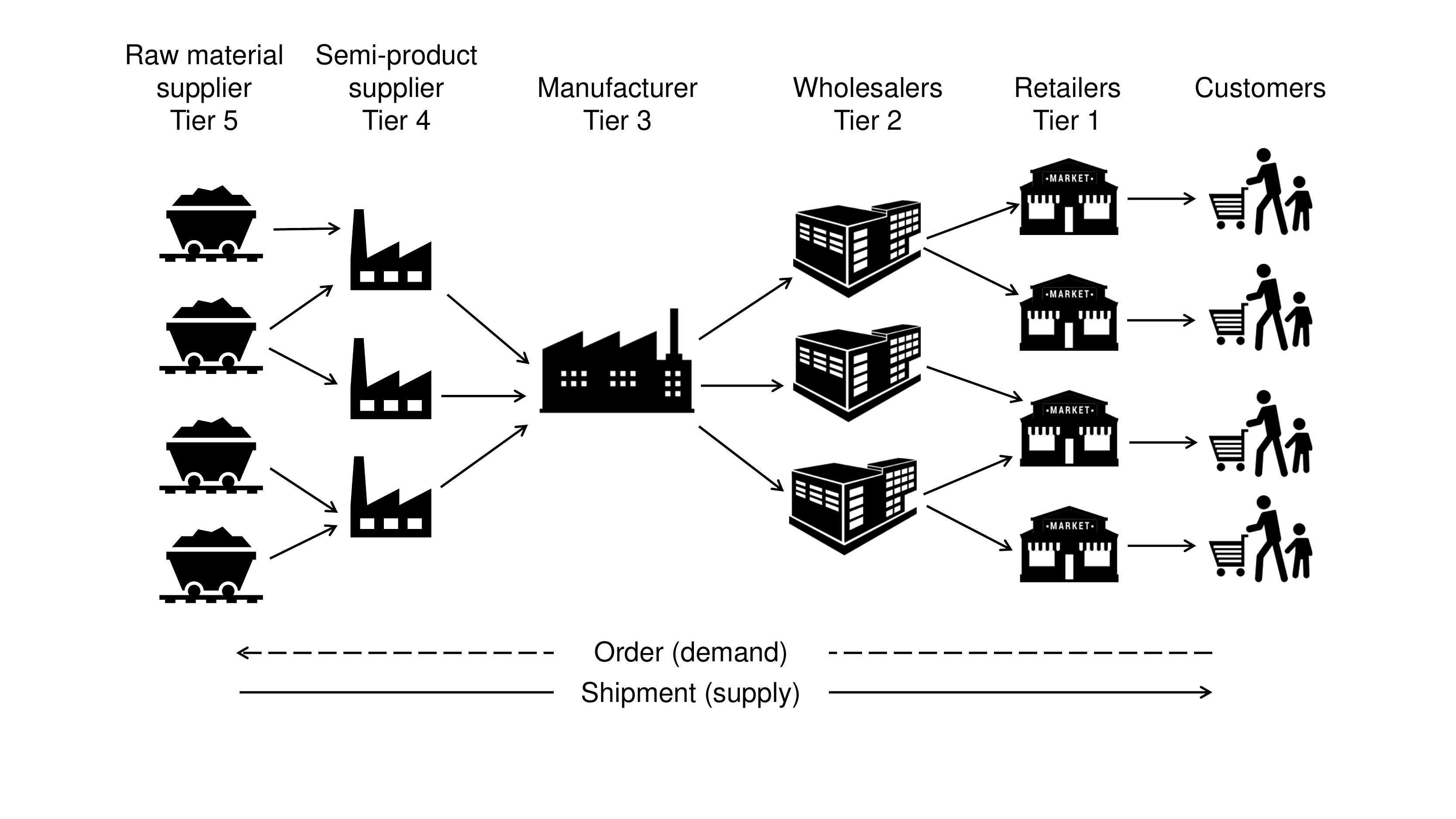}
\caption{Schematic of a supply network}
\label{fig:supply_net_eg}
\end{figure}

In this study, we seek to investigate the BWE of general supply networks. In the existing literature on the BWE of supply chains, the majority of studies have focused on serial structure~\citep{ouyang2010bullwhip,brintrup2018supply,giri2022bullwhip} due to (i) the scarcity of empirical data about the topology as a result of firms' confidentiality concerns and (ii) challenges in the computation and modeling of large-scale supply networks. However, real supply networks are rarely a single chain, as there can be many suppliers in the same layer of the supply chain for a single product and manufacturers of similar products (e.g., automobiles) can share multiple suppliers upstream and buyers downstream~\citep{brintrup2018supply} (Layer 5 in Fig. \ref{fig:supply_net_eg}). As such, it is critical to characterize the BWE of complex supply networks with more complex structures. Additionally, the supply chains of firms typically have a few layers only and each layer can have a large number of suppliers~\citep{willems2008data,kito2014structure,bode2015structural}, thus it is important to characterize how the (non)stationarity of order over a few layers upward from the market impact the BWE.

The contributions of our research to the extensive literature on BWE are four-fold: \textbf{First}, we derive the model for quantifying the layer-wise BWE of general supply networks, rather than a serial supply chain using the control-theoretic approach that enables the analysis in the frequency domain \citep{dejonckheere2003measuring}. The inventory replenishment policy is the popular order-up-to policy in industry. \textbf{Second}, we analytically and numerically show that when subject to market demand generated from the same stationary process, the structure of supply networks does not affect the layer-wise BWE of the supply networks. The layer-wise BWE of supply networks of different structures is equivalent to that of the corresponding serial network. \textbf{Third}, we show that wider supply networks have lower layer-wise BWE when subject to market demand generated from different stationary and non-stationary processes. \textbf{Fourth}, we provide a characterization of the node-to-node BWE of general supply networks when nodes do not have a unique layer position due to the presence of intra-layer links between nodes or inter-layer links between nodes that are not positioned in consecutive layers.

The remainder of this paper is structured as follows. Section \ref{sec:lit_review} presents the literature review. In Section \ref{sec:meth}, we introduce the inventory management model as well as the metric. We also formulate the problem and provide analytical derivations and propositions. Section \ref{sec:result} uses numerical experiments to demonstrate the methods and validate the propositions. Conclusions and a discussion of possible extensions of this work are provided in Section \ref{sec:conc}. Long proofs are included in the Appendix.

\section{Literature Review} \label{sec:lit_review}

First discovered by Procter \& Gamble (P\&G) and then brought to extensive academic attention by \citet{lee1997bullwhip}, BWE of supply chains has been explored extensively \citep{chen2000quantifying,ouyang2006characterization,cachon2007search,ouyang2010bullwhip,dominguez2014serial,osadchiy2021bullwhip}. Because the literature on the analysis of BWE is vast, we confine our attention to studies that focus on the analytical characterization of BWE as well as the impact of market demand or structure on BWE. For a more comprehensive review on BWE, readers are referred to \citet{wang2016bullwhip}.

There are two main approaches to analytically characterize the BWE. The first approach is statistical, which is to derive the variance of orders made by suppliers in the time domain. After the seminal work by \citet{lee1997bullwhip}, this approach has been widely adopted to investigate BWE under influential factors, such as lead time, information sharing, batching, and operational deviations~\citep{chen2000quantifying,gaur2005information,ma2013bullwhip,chen2017measuring,sodhi2011incremental}. However, to maintain analytical tractability in deriving the variance of orders in the time domain, most studies in this line of research have focused on simple serial supply networks only. The focus on serial structure may limit the generality of those studies because most supply chains have complex structures due to complex relations among suppliers. One exception is \citep{sodhi2011incremental}. The authors examine the incremental BWE of an arborescent (treelike) supply chain due to the operational deviations, such as errors in order placement and lag in sharing demand prediction. They demonstrate that even without operational deviations, the BWE for a node (supplier) in a tree-like supply network subject to stationary market demand (autoregressive process) is exacerbated as the number of layers, the number of downstream suppliers, or the lead times increases. However, non-stationary market demand is not considered. The other exception is \citet{dominguez2014serial}, in which the authors combine statistical and numerical analyses to compare the BWEs (order variance ratio) of 4-layer serial and divergent supply networks. Their simulation results show that (i) when subject to normally distributed demand, the BWE for serial and divergent supply networks are the same, while (ii) when subject to stochastic step demand (the mean of the normal distribution jumps to a much higher value), the divergent network leads to higher BWE than serial network. However, the market demand is an oversimplification and an analytical characterization of how the two structures lead to different BWEs is not provided.


The second approach is control-theoretical, which is to quantify the BWE of the supply chain in the frequency domain. First introduced by \citet{dejonckheere2003measuring}, this approach employs the $\cZ$ transformation to derive the BWE based upon the transfer function that maps the output (order by a supplier) to the input (order by the downstream supplier/customer). The $\cZ$-transformation, a discrete analogue of the Laplace transform, is used due to the discrete-time models of inventory management and (ii) more tractable characterization of BWE, especially for multi-echelon supply chains. In particular,
\citet{dejonckheere2003measuring} use this approach to characterize the BWE of a single echelon supply chain. \citet{jakvsivc2008effect} apply this approach to investigate the impact of generalized order-up-to policies on the BWE of a single echelon. Adopting the same control theoretic approach, \citet{ouyang2006characterization} establish the condition for the existence of BWE (worse-case amplification rate of the standard deviation of orders) and derive the layer-wise transfer function to characterize the BWE in serial supply chains. Later, \citet{ouyang2010bullwhip} extend this approach to characterize the systemic BWE of supply networks.

In our work, the methodology is mainly built upon the control-theoretic approach, but our work differs from existing related work on characterizing the BWE of supply chains because we focus on the BWE of general supply networks instead of serial or tree-like supply chains and then investigate the joint impact of network structure and market demand on the BWE. We also characterize the BWE of supply networks when the layer position of a node might be nonunique because of intra-layer links among nodes in supply networks. To the best of our knowledge, our work represents the first step towards this direction of research.

\section{Methodology} \label{sec:meth}
\subsection{Inventory Control Policy}

We adopt the order-up-to replenishment policy because order-up-to is one of the most popular inventory replenishment policies in industry and following \citet{lee1997bullwhip}, a large number of existing articles adopt this policy as well \citep{chatfield2004bullwhip,luong2007measure,dominguez2014serial,nagaraja2015measuring}. The sequence of actions in inventory management under this policy are as follows:

\begin{enumerate} \itemsep0pt 
    \item At the beginning of each period, update the order-up-to level using the forecasted demand.
    \item Place an order to raise or lower the inventory position to the order-up-to level.
    \item After the lead time, the supply of goods from the suppliers (upstream nodes) is received.
    \item Receive new orders from downstream demand nodes and satisfies demand.
    \item Calculate a new forecasted demand using the demand from previous periods to be used in the next period.
\end{enumerate}
The model involves multiple rules: (i) negative orders (i.e., returns) are allowed, (ii) placed orders will not be canceled, and (iii) the inventory position is allowed to be negative. 
Note that each firm only needs to determine the quantity of goods to order at the beginning of its review time period. In this paper, we assume that suppliers use a simple moving average to forecast demand. The lead time is the time gap between order placement and delivery. The minimum value of the lead time is 1, in which case the order is received at the next time period of order placement.

Based on the order-up-to replenishment policy, the dynamics of supply networks are presented as follows. Consider a general supply network $\cG=\left(\cV, \cE\right)$ wherein the suppliers (nodes) are presented by the set $\cV$ and the directed connections among suppliers (links) are represented by the set $\cE$. The dynamics of a supply network can then be expressed as

\vspace{-8pt}
\begin{subequations}\label{eq:dynamics_net}
\begin{align} 
& \label{eq:node_dynamics_net}
    {x_i}\left( {t + 1} \right) - {x_i}\left( t \right) = \underbrace {\sum\limits_{k = 1}^{{\left|\cV\right|}} {{\bA_{ik}}} {{y} _{ik}}\left( t \right)}_{{\rm{Total~supply}}} - \underbrace {\sum\limits_{k = 1}^{{\left|\cV\right|}} {{\bA_{ki}}} {{y} _{ki}}\left( t \right)}_{{\rm{Total~ demand}}},\;\forall i \in \cV,\\
& \label{eq:link_dynamics_net}
    {{y} _{ij}}\left( t \right) = \frac{{{\bA_{ij}}}}{{\sum\limits_{k = 1}^{{\left|\cV\right|}} {{\bA_{ik}}} }}\left( {\underbrace { - {x_i}\left( t \right) + \underbrace {{L_i} \cdot \frac{{\sum\limits_{k = 1}^{{\left|\cV\right|}} {{\bA_{ki}}} \left[ {{{y} _{ki}}\left( {t - 1} \right) + \ldots + {{y} _{ki}}\left( {t - P_i} \right)} \right]}}{{P_i}}}_{{\rm{Forecasted~ demand}}}}_{{\rm{Order - up - to~level}}}} \right), \;
    \forall \left( {i,j} \right) \in \cE,
\end{align}
\end{subequations}

\noindent
where $x_i$ represents the inventory position of supplier $i$. ${y}_{ij}$ represents the quantity of orders from supplier $i$ to supplier $j$. $A$ is the adjacency matrix of the supply networks: $\bA_{ij}=1$ indicates the presence of link from supplier $i$ to supplier $j$, and 0 otherwise. Note that since there are no self links, $\bA_{ii}=0$ always holds. $L_i$ and $P_i$ are the lead time of supplier $i$ and the number of previous time periods used in estimating the moving average demand. $(i,j)$ is the link from node $i$ to node $j$. ${\left|\cV\right|}$ is the number of nodes in the supply network.

\begin{table}[!b]
\begin{tabular}{lp{14.3cm}l}
\midrule
\multicolumn{2}{c}{\textbf{Notations}} \\
\multicolumn{2}{l}{\textbf{Sets \& Indices}} \\
$\cV$, $\cE$, $\cL$ & Sets of nodes, links, and layers\\
$i,\;j$ & Indices of nodes\\
$(i,j)$ & Index of links\\
${l}$ & Index of layers\\
$t$ & Index of time periods\\
\multicolumn{2}{l}{\textbf{Variables \& Parameters}} \\
$a$ & Slope of trend for market demand over time\\
$c$ & Constant in the model for simulating market demand\\
$h$  & Number of order seasonality components in market demand \\
$x_i$ & Inventory position of node $i$\\
$y_{0}$ & Order quantity (demand) from market\\
$y_{ij}$ & Order quantity from node $i$ to $j$\\
$\cZ$ & $\cZ$-transformation operator that converts a sequence of numbers into the $z$-domain\\
$\bm A$ & Adjacency matrix\\
$\bm W$ & Weight matrix\\
$\bm B$ & Matrix for the amplification rate from non-absorbing nodes to absorbing nodes \\
$M$ & Number of layer width (number of input demand sequences)\\
$N$ & Number of generic numbers\\
$N_a$, $N_t$ & Number of absorbing nodes, number of non-absorbing nodes\\
$\bm Q$ & Matrix for the connectivity from non-absorbing nodes to absorbing nodes\\
$\bm R$ & Matrix for the connectivity among non-absorbing nodes\\
$L_i$, $P_i$  & Lead time and number of time periods used in estimating average demand from supplier $i$ \\
$T$ & Number of time periods\\
$Y_{ij}$ & Sequence of order quantity from $i$ to $j$\\
$\gamma$, $v$ & Amplitude and frequency for seasonality component of market demand\\
$\omega$ & Generic frequency\\
$\epsilon$, $\sigma$ & Noise term and the standard deviation of noise\\
$\cA$ & Amplitude\\
$\cF_i$ & Transfer function for node $i$\\
$\phi$ &  Amplification rate of amplitude at a frequency\\
$\varphi$ & Parameter in the AR(1) process\\
$\Phi$ &  Layer-wise BWE\\
\midrule
\end{tabular}
\end{table}

For serial structure, the dynamics given by Eq. \eqref{eq:dynamics_net} reduce to Eq. \eqref{eq:dynamics_chain}. Note that $y_{01}$ ($i=0$) represents the input market demand.

\vspace{-6pt}
\begin{subequations} \label{eq:dynamics_chain}
\begin{align} 
    & \label{eq:node_dynamics_chain} {x_i}\left( {t + 1} \right) - {x_i}\left( t \right) = {y_{i, i+1 }}\left( t \right) - {y_{(i - 1),i}}\left( t \right),\;\forall i \in \cV, \\
    & \label{eq:link_dynamics_chain}
    {y_{i, i+1}}\left( t \right) =  - {x_i}\left( t \right) + {L_i} \cdot \frac{{{y_{i-1,i}}\left( {t - 1} \right) + \ldots + {y_{i-1,i}}\left( {t - P_i} \right)}}{{P_i}},\;\forall i \in \cV.
\end{align}
\end{subequations}

\noindent Plugging Eq. \eqref{eq:link_dynamics_chain} into Eq. \eqref{eq:node_dynamics_chain}, we can simplify the dynamics for linear networks at a node as

\vspace{-6pt}
\begin{subequations} 
\begin{align}
    &\label{eq:node_dynamics_chain_simple}
    {x_i}\left( {t + 1} \right) = {L_i} \cdot \frac{{{y_{i-1,i}}\left( {t - 1} \right) + \ldots + {y_{i-1,i}}\left( {t - P_i} \right)}}{{P_i}} - {y_{i-1,i}}\left( t \right),\;\forall i \in \cV,\\
   &\label{eq:link_dynamics_chain_simple}
   {y_{i,i+1}}\left( {t} \right) =  {L_i} \cdot \frac{{{y_{i-1,i}}\left( t-1 \right) - {y_{i-1,i}}\left( {t - P_i-1} \right)}}{{P_i}} + {y_{i-1,i}}\left( t - 1\right),\;\forall i \in \cV.
\end{align}
\end{subequations}


\subsection{Metric for the Bullwhip Effect}

A commonly used metric for BWE in serial supply chain is the ratio of the standard deviation of order from a node and the order into the node \citep{dejonckheere2003measuring,dejonckheere2004impact,jakvsivc2008effect}. Formally, BWE for a node in serial supply chain is

\vspace{-6pt}
\begin{equation}
 \phi_i  = \sqrt{\frac{\var\left(y_{i,i+1}\right)}{\var\left(y_{i-1,i}\right)} },\; \forall \, i \in \cV.
\end{equation}

\noindent
To simplify the notations, we use $\var\left( y_{ij} \right)$ to represent $\var\left(\{y_{ij}\}^T_0 \right)$ in calculating the variance where $T$ is the number of time periods. For supply networks, this metric can be extended as

\vspace{-6pt}
\begin{equation}
    \phi_i = \sqrt{\frac{\textrm{var} \left( {\sum\limits_{j = 1}^{{\left|\cV\right|}} {{\bA_{ij}}} {{Y} _{ij}}}\right)} {{\textrm{var} \left( {\sum\limits_{k = 1}^{{\left|\cV\right|}} {{\bA_{ki}}} {{Y} _{ki}}}\right)}}}, \; \forall \, i \in \cV,
\end{equation}
\noindent
where $Y_{ij} = \{y_{ij}(t)\}_{1}^{T}$ is the sequence of order quantity from $i$ to $j$.


Next we define the layer-wise BWE for supply networks whose nodes have single layer position. The layer in which a supplier is located is defined as the length of the shortest path from a supplier to a customer. For example, all retailers are located in layer one. Layer-wise BWE of layer ${l}$, denoted as $\Phi_l$, is defined as the ratio of the standard deviation of total order from a layer and the standard deviation of total order to the layer, i.e.,

\vspace{-6pt}
\begin{equation}
    \Phi_{{l}} = \sqrt{\frac{\textrm{var} \left( \sum_{i \in \cL_{{l}}} \sum\limits_{j=1}^{|\cV|} {\bA}_{ij} {y}_{ij}\right)} {{\textrm{var} \left( \sum\limits_{k=1}^{|\cV|}\sum_{i \in \cL_{{l}}} {\bA}_{ki} {y}_{ki}\right)}}}, \; \forall \, {l} \in \cL,
\end{equation}

\noindent
where $\cL$ is the set of layers of a supply network.

\subsection{Frequency domain analysis}

Order sequences from customers and suppliers in the supply networks that show demand change in the time domain can be represented equivalently in the frequency domain. The frequency domain translation of time domain data can allow for more convenient and simpler analysis of system output with respect to the input. For stationary sequences, the $\cZ$-transformation is commonly employed~\citep{dejonckheere2003measuring,ouyang2006characterization,jakvsivc2008effect}. The $\cZ$-transformation (unilateral) of a given sequence of numbers $\{ \bm{x} _n \}_{0}^{\infty} $ is conducted according to
$X \left[z \right] = \sum_{n=0}^{\infty} x_n z^{-n}$. If we substitute $e^{i \omega}$ for $z$ where $i=\sqrt{-1}$, then the $\cZ$-transformation reduces to the discrete Fourier transformation (DFT)~\citep{mitra2006digital} and the transfer function is converted to the frequency domain. In particular, the $\cZ$-transformation has the following two properties that are useful in deriving the transfer function corresponding to the supply network dynamics in Eq.~\eqref{eq:dynamics_net}: (i) $\cZ \left[x (t) + y(t) \right] = \cX + \cY$ where $\cX$ and $\cY$ are the $\cZ$-transformation of $x$ and $y$ (linearity) and (ii) $\cZ \left[x(t-k)\right] = z^{-k} \cX$ as well as $\cZ \left[x(t+1)\right] = z \cX - zx(0)$ (time shift) \citep{baraniuk2009signals}. Note that the $\cZ$-transformation, similar to the Fourier transformation, is typically applicable to stationary (periodic) sequence~\citep{osgood2002lecture}, thus it is not suitable for non-stationary sequences whose frequency components can vary over time. Applying the $\cZ$-transformation to Eq.~\eqref{eq:dynamics_net} and assuming that $x_i(0)=0$ (initial inventory level of all suppliers are zero), we can get the dynamics of supply networks in the $\cZ$ space

\vspace{-6pt}
\begin{subequations}\label{eq:dynamics_net_z}
\begin{align} 
& \label{eq:node_dynamics_net_z}
    \left(z-1\right) {\cX _i} = \sum\limits_{j = 1}^{\left|\cV\right|} {   {\bm{A}_{ij}}} {\cY_{ij}} - \sum\limits_{k = 1}^{\left|\cV\right|} {   {\bm{A}_{ki}}} {\cY_{ki}},\;\forall i \in \cV,\\
& \label{eq:link_dynamics_net_z}
    {\cY _{ij}}     = \frac{{{\bA_{ij}}}}{{\sum\limits_{k = 1}^{{\left|\cV\right|}} {{\bA_{ik}}} }} \left(  - {\cX _i} + \frac{{{L_i}\left( {{z^{ - 1}} +  \ldots  + {z^{ - {P_i}}}} \right)\sum\limits_{k = 1}^{\left|\cV\right|} {{A_{ki}}} {\cY_{ki}}}}{{{P_i}}}  \right), \;
    \forall \left( {i,j} \right) \in \cE.
\end{align}
\end{subequations}

Using Eq. \eqref{eq:node_dynamics_net_z} in Eq. \eqref{eq:link_dynamics_net_z} and some algebraic manipulations, we can get the transfer rate (function) of the output from and the input into a node, which is given by

\vspace{-6pt}
\begin{equation} \label{eq:node_transfer_func}
   \cF_i (z) = \frac{{\sum\limits_{k = 1}^{\left|\cV\right|} {{A_{ik}}} {\cY_{ik}}}}{{\sum\limits_{k = 1}^{\left|\cV\right|} {{A_{ki}}{\cY_{ki}}} }} 
   =\frac{{1 + \frac{{{L_i}}}{{{P_i}}}\left( {1 - {z^{ - {P_i}}}} \right)}}{z}, \; \forall\, i \in \cV.
\end{equation}
\noindent
If the demand into nodes in the same layer of a supply networks are independent, then layer-wise transfer function is the sum of the transfer rates for all nodes in the same layer

\vspace{-6pt}
\begin{equation} \label{eq:layer_transfer_func}
    \cF_l (z) = \sum_{i \in \cL_l}{\cF_i (z)} = \sum_{i \in \cL_l}{ \frac{{1 + \frac{{{L_i}}}{{{P_i}}}\left( {1 - {z^{ - {P_i}}}} \right)}}{z}}, \; \forall\, l \in \cL.
\end{equation}





\subsubsection{Layer-wise BWE}
In the frequency domain, the layer-wise BWE with linear dynamics (Eq. \eqref{eq:dynamics_net}) is determined by how the amplitude of the component for each frequency changes along the supply networks. This is because as the demand signal propagates along the supply network, the frequency of a sinusoidal wave will be the same, while its amplitude and phase can change. Since the standard deviation of a sinusoidal wave can be calculated by its amplitude according to $\var(s) = \frac{\cA^2}{2}$~\citep{lyons2004understanding},
the variance amplification of stationary demand from a single echelon (node) can be measured according to~\citep{dejonckheere2003measuring}

\vspace{-6pt}
\begin{equation}
   \phi^{(1)} = 
   \sqrt {\frac{{\sum\limits_{n = 1}^{T/2 - 1} {\cA_n^2} \phi^2_{ _n} }}{{\sum\limits_{n = 1}^{T/2 - 1} {\cA_n^2} }}}, \; \forall i \in \cV,    
\end{equation}

\noindent
where $\phi_n$ is the amplification rate of the demand signal component with frequency $\omega_n$ obtained using the Fourier transformation of the demand signal into node $i$. The amplitude $\mathcal A$ of each frequency for the market demand component can be obtained by DFT. 
The DFT of a given sequence of $N$ numbers $ \{ \bm{x} _n \}_{0}^{N-1} $ to $\left\{\bm {X} _{k}\right\}_{0}^{N-1}$ according to 
$X_k := \sum_{n=0}^{N-1} x_n \, e^{- 2 \pi \, i \,  k n/N}$.
Given Eq.~\eqref{eq:node_transfer_func} and the fact that the magnitude of the transfer rate is $\left| \cF (e^{i\omega_n}) \right|^2 = \cF (e^{i\omega_n}) \cdot \cF (e^{-i\omega_n})$ in the frequency domain, we get

\vspace{-6pt}
\begin{equation} \label{eq:amp_amplif_by_freq}
    \phi^2_i\left( \omega \right) 
    = \frac{{1 + \frac{{{L_i}}}{{{P_i}}}\left( {1 - {e^{ - {P_i}i\omega }}} \right)}}{{{e^{i\omega }}}} \cdot \frac{{1 + \frac{{{L_i}}}{{{P_i}}}\left( {1 - {e^{{P_i}i\omega }}} \right)}}{{{e^{ - i\omega }}}} = 1 + 2\left( {\frac{{{L_i}}}{{{P_i}}} + \frac{{L_i^2}}{{P_i^2}}} \right)\left( {1 - \cos \left( {{P_i}\omega } \right)} \right) .
\end{equation}
\noindent
Since $0 \le {1 - \cos \left( {{P_i}\omega } \right)} \le 1$, $\forall \omega \ge 0,\;\phi_i\left( \omega \right) \ge 1$. Therefore, under the order-up-to policy, the bullwhip effect will occur in a single echelon supply chain.

\begin{figure} [htb]
    \centering
    \includegraphics[width=0.8\textwidth]{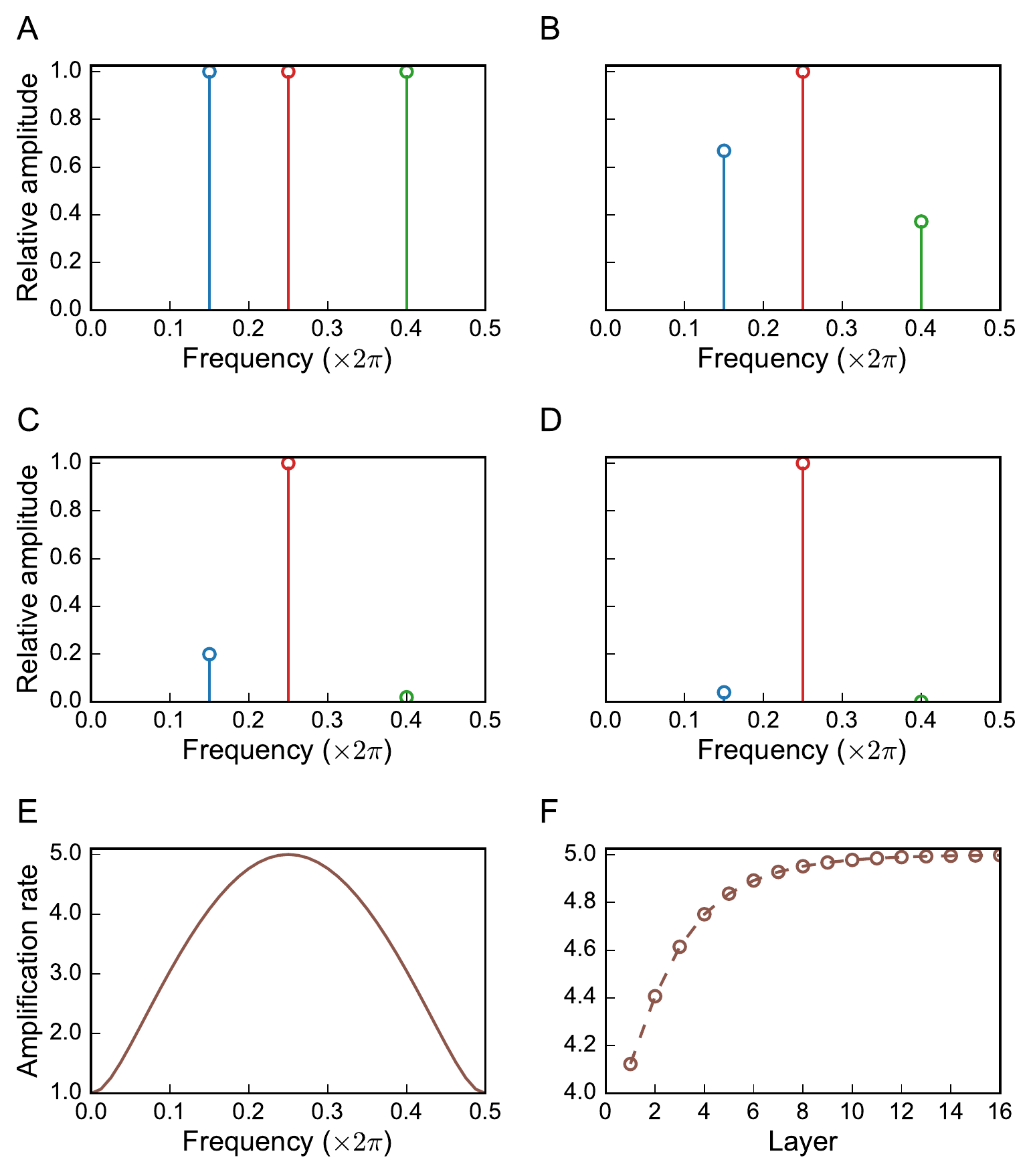}
    \caption{Amplification of amplitudes of different frequencies along a serial supply network. A, B, C, and D correspond to the relative amplitude (to the maximum amplitude) for each frequency in demands from the market, layers 2,  8, and 16, respectively. The market demand is simulated using $ y_{t} = \sin(2\pi\cdot 0.15 t) +  \sin(2 \pi \cdot 0.25 t) + \sin(2 \pi \cdot 0.40 t)$. The moving average window in estimating the order-up-to level is 2 while lead time is 4. E. The amplification rate for different frequencies. Among the three frequencies of sinusoidal market demand, frequency $0.25\cdot2\pi$ has the highest amplification rate, followed by that with frequency $0.15\cdot2\pi$ and then $0.40\cdot2\pi$. F. The respective amplification rate for each layer.}
    \label{fig:amp_amplif}
    \vspace{-6pt}
\end{figure}

Using DFT, each stationary component of the market demand sequence can be decomposed into the sum of $T/2-1$ sinusoidal waves with different amplitudes and frequencies. Note that for non-stationary market demand, the standard deviation can no longer be directly calculated the amplitudes of frequencies using DFT. The variance of a stationary sequence of market demand $\left\{ y_{0m}\right\}$ after going through layer ${l}$ is

\vspace{-6pt}
\begin{equation} \label{eq:layerwise_var}
    \var \left( y_{{l} m} \right) = {\frac{1}{2}\sum\limits_{n = 1}^{T/2 - 1} {{\phi ^{2l}_n}{\cA}_{mn}^2} },\; \forall \; {l} \in \cL,
\end{equation}
\noindent
where ${\cA}_{mn}$ is the amplitude of the frequency $\omega_n$ for $\left\{ y_{0m}\right\}$. It is obvious that a higher lead time will lead to higher amplification rate and therefore a higher order variance. Then, the layer-wise BWE in layer ${l}$ of a supply network is given by

\vspace{-6pt}
\begin{equation} \label{eq:layerwise_bwe}
    {\Phi_{{l}}} = \sqrt {\frac{{\sum\limits_{n = 1}^{T/2-1} {{{ \phi }^{2l}_n}\cA_{n}^2       } }}{{\sum\limits_{n = 1}^{T/2-1} {{{ \phi  }^{2 \left({l} - 1\right)}_n}\cA_{n}^2       } }}},\; \forall \;{l} \in \cL.
\end{equation}



From Eq. \eqref{eq:layerwise_bwe}, we can see that BWE occurs because as the demand propagates upstream according to the system dynamics (in this case the order-up-to policy), the amplitude of the demand component whose frequency leads to higher amplification rate become increasingly dominate, such as the amplitude for frequency $0.25\cdot2\pi$ in Fig. \ref{fig:amp_amplif}. The value of largest BWE is determined by the demand component with the dominant frequency.

Given Eqs. \eqref{eq:amp_amplif_by_freq} and \eqref{eq:layerwise_bwe}, we can further achieve the following results on the relation between tier-wise BWE and the layer position (Proposition \ref{thm:max_BWE}) and the impact of lead time on layer-wise BWE (Proposition \ref{thm:lead_time_BWE}).

\begin{proposition} \label{thm:max_BWE}
In a supply network with dynamics defined by Eq. \eqref{eq:dynamics_net}, if the lead times and time window lengths are equal for all nodes, then (i) the layer-wise BWE monotonically increases with $l$ and (ii) the limit of the layer-wise BWE equals to the maximum value of the amplification rate among all frequencies given by Eq \eqref{eq:amp_amplif_by_freq}.
\end{proposition}

\begin{proof} [Proof] 
Using Eq. \eqref{eq:layerwise_bwe}, we have

\vspace{-6pt}
\begin{equation}
    \frac{ {\Phi _l^2} } {{ {\Phi _{l-1}^2} } } = { {\frac{{\sum\limits_{n = 1}^{T/2 - 1} {{\cA}_n^2\phi _n^{2l}} }}{{\sum\limits_{n = 1}^{T/2 - 1} {{\cA}_n^2 \phi _n^{2l-2}} }}}} \cdot { {\frac{{\sum\limits_{n = 1}^{T/2 - 1} {{\cA}_n^2\phi _n^{2l-4}} }}{{\sum\limits_{n = 1}^{T/2 - 1} {{\cA}_n^2\phi _n^{2l-2}} }}}}.
\end{equation}

\noindent
By Cauchy-Schwartz inequality, for $\phi_n\neq 1$ we observe
\begin{equation}
   \sum_{n=1}^{T/2-1}\cA_n^2\phi_n^{2l}\cdot\sum_{n=1}^{T/2-1}\cA_n^2\phi_n^{2l-4}> \left(\sum_{n=1}^{T/2-1}\cA_n^2\phi_n^{2l-2}\right)^2, 
\end{equation}
\noindent
therefore $\Phi_l^2 > \Phi_{l-1}^2$. Because $\Phi_l > 0$ and $\Phi_{l-1}>0$, $\Phi_l > \Phi_{l-1}$.

Next we prove the second part. Suppose the maximum value of amplification rate is achieved at $\omega_k$, i.e., $\phi_k= \phi_{\rm{max}}$, therefore $\frac{{\phi _n^2}}{{\phi _{k }^2}} < 1, \; \forall\, n \ne k$ and then 

\vspace{-6pt}
\begin{equation}
\mathop {\lim }\limits_{l \to \infty } {\Phi _l} = \mathop {\lim }\limits_{l \to \infty } \sqrt {\frac{{\sum\limits_{n = 1}^{T/2 - 1} {\frac{{\phi _n^{2l}}}{{\phi _k^{2l}}}{\cA}_n^2} }}{{\sum\limits_{n = 1}^{T/2 - 1} {\frac{{\phi _n^{2\left( {l - 1} \right)}}}{{\phi _k^{2l}}}{\cA}_n^2} }}} = \sqrt {\mathop {\lim }\limits_{l \to \infty } \frac{{\sum\limits_{n = 1}^{T/2 - 1} {\frac{{\phi _n^{2l}}}{{\phi _k^{2l}}}{\cA}_n^2} }}{{\sum\limits_{n = 1}^{T/2 - 1} {\frac{{\phi _n^{2\left( {l - 1} \right)}}}{{\phi _k^{2l}}}{\cA}_n^2} }}}  = \sqrt {\mathop {\lim }\limits_{l \to \infty } \frac{{\frac{{\phi _k^{2l}}}{{\phi _k^{2l}}}{\cA}_k^2}}{{\frac{{\phi _k^{2\left( {l - 1} \right)}}}{{\phi _k^{2l}}}{\cA}_k^2}}}  = \sqrt {\phi _k^2}  = {\phi _{\max }}.    
\end{equation}
\end{proof}

\begin{proposition} \label{thm:lead_time_BWE}
In a supply network with dynamics defined by Eq. \eqref{eq:dynamics_net}, a higher replenishment lead time in a layer or its downstream layer does not always lead to higher layer-wise BWE.
\end{proposition}

\begin{proof} [Proof] 
Since $\phi$ is a monotonically increasing function of lead time, we consider the partial derivative of $\Phi_l^2$ with respect to $\phi$. 
Let $G = \sum\limits_{n = 1}^{T/2 - 1} {{\cA}_n^2\phi _n^{2(l - 1)}}$, then we have

\vspace{-6pt}

\begin{subequations}
\begin{align}
\frac{\partial \Phi_l^2}{\partial \phi} 
& = {{{G^{-2}}}} \left({\sum\limits_{n = 1}^{T/2 - 1} {{\cA}_n^22l\phi _n^{ - 1}\phi _n^{2l} \cdot \sum\limits_{n = 1}^{T/2 - 1} {{\cA}_n^2\phi _n^{ - 2}\phi _n^{2l}} }  - \sum\limits_{n = 1}^{T/2 - 1} {{\cA}_n^2\phi _n^{2l}}  \cdot \sum\limits_{n = 1}^{T/2 - 1} {{\cA}_n^22(l - 1)\phi _n^{ - 3}\phi _n^{2l}} } \right)\\
& = 2 {{{G^{-2}}}} \left({\sum\limits_{n = 1}^{T/2 - 1} {\sum\limits_{m = 1}^{T/2 - 1} {{\cA}_m^2 {\cA}_n^2\phi _m^{2l}} \phi _n^{2l} \cdot l\phi _m^{ - 2}\phi _n^{ - 1}}  - \sum\limits_{n = 1}^{T/2 - 1} {\sum\limits_{m = 1}^{T/2 - 1} {{\cA}_m^2 {\cA}_n^2} \phi _m^{2l}\phi _n^{2l} \cdot (l - 1)\phi _m^{ - 3}} }\right)\\
& = 2 {{{G^{-2}}}} {\sum\limits_{n = 1}^{T/2 - 1} {\sum\limits_{m = 1}^{T/2 - 1} {{\cA}_m^2 {\cA}_n^2\phi _m^{2l - 1}\phi _n^{2l-1} \cdot \left( {l - (l - 1) \phi _n \phi _m^{ - 1}} \right)} } }  \label{eq:phi_deriv_L_last}
\end{align}
\end{subequations}



\noindent
We can see that $\frac{\partial \Phi_l^2}{\partial \phi} \ge 0$ is always true only when $ \phi _n \phi _m^{ - 1} (l-1) \le l, \; \forall m,\,n \in \{1,2,...,T/2\}$. Therefore, we can observe that except for the trivial case where $l=1$, $\frac{\partial \Phi_l^2}{\partial \phi} \ge 0$ does not always hold when $l \ge 2$.
\end{proof}

\begin{rmk} \label{rmk:BWE_vs_leadtime}
We note that this result does not agree with the implication of Theorem 3 in \citet{sodhi2011incremental} for an arborescent supply chain (a node at level $l$ have $l$ downstream nodes), which is established using the statistical approach. The reason for the difference is that in their result, the bullwhip effect is quantified by the order variance instead of the order variance ratio.
\end{rmk}

\subsubsection{Node-to-node BWE}

The layer-wise BWE defined above is not applicable when the supply network has suppliers whose layer position is not unique, such as when there exist intra-layer links or links between suppliers that are not positioned in consecutive layers. In this case, we used the node-to-node amplification rate to examine the BWE. The node-to-node amplification rate is a measure of how much the fluctuations of market demand input into a node $i$ is amplified when it moves upstream to one of the final, external suppliers, $k$. Formally, the amplification rate is defined as

\vspace{-6pt}
\begin{equation}
    \Phi_{{k \leftarrow i}}=\sqrt{\frac{\var\left( y_{{k \leftarrow i}}\right)}{\var \left(y_{0i}\right)}}.
\end{equation}

To derive the formula for $\Phi_{{k \leftarrow i}}$, we borrow the idea of absorbing Markov Chain \citep{grinstead1997introduction}, in which the absorbing node (state) is a node that cannot be left once entered. In the supply network, external suppliers, which are connected to the uppermost nodes in the supply networks (Fig. \ref{fig:structure_types}), are treated as the absorbing nodes. The rest of the supply nodes are treated as non-absorbing nodes. The transition matrix in the absorbing Markov chain captures the probabilities of transitioning from one node to another in one step. We adopt this matrix to encode the connectivity among non-absorbing and absorbing nodes, which is given by 




\vspace{-6pt}
\begin{equation}
  {\bm H }=\left[\begin{array}{ll}
{\bm I}_{a} & \bm 0 \\
\bm R & \bm{W}
\end{array}\right]  ,
\end{equation}

\noindent
where ${\bm I}_{a}$ is an identity matrix of dimension $N_a$, which represents the number of absorbing nodes, i.e., suppliers without upstream suppliers. $\bm R^{ N_t\times N_a}$ is a matrix containing the weights of links from non-absorbing nodes to the absorbing nodes wherein $N_t$ representing the numbers of non-absorbing nodes, i.e., suppliers with upstream suppliers. $\bm W^{ N_t\times N_t}$ is the matrix containing the weights of links from non-absorbing nodes to non-absorbing nodes, i.e., the weight matrix of the original network that determines how the order from a node is divided among its upstream suppliers . The total amplification rate of demand component with frequency $\omega_n$ of a demand sequence after going through all non-absorbing nodes can be given by 

\vspace{-6pt}
\begin{equation} \label{eq:b_orig}
\bm B \left(\omega_n\right) = \left(\bm I_{t}- \bm W \odot \bm{\phi} \left(\omega_n\right)\right)^{-1} \bm R \odot \bm{\phi} \left(\omega_n\right),
\end{equation}
\noindent

\noindent
where ${\bm I}_{t}$ is an identify matrix of dimension $N_t$. $\odot$ is the operator for element-wise multiplication. $\bm{\phi}\left(\omega_n\right)$ is the matrix for the amplification rate at each non-absorbing node for the demand component with frequency $\omega_n$. The amplitude of the market demand component with frequency $\omega_n$ to the absorbing node $k$ is the first row in the respective matrix $\bm B$, denoted by $\bm B_{1k} (\omega_n)$. For homogeneous supply networks, $\bm{\phi} \left(\omega_n\right)$ is reduced to a scalar $\phi\left(\omega_n\right)$, then Eq. \eqref{eq:b_orig} reduces to

\vspace{-6pt}
\begin{equation} \label{eq:b_orig_homo}
\bm B \left(\omega_n\right) = \left(\bm I_{t} \phi^{-1} \left(\omega_n\right) - \bm W \right)^{-1} \bm R.
\end{equation}

\noindent
Let $\cA_{in}$ represent the amplitude of the frequency $\omega_n$ for market demand sequence generated from node $i$, and the variance of demand component at absorbing node $k$ from demand sequence generated from node $i$ can be given by

\vspace{-6pt}
\begin{equation}
    {\mathop{\rm var}} \left( {{y_{k \leftarrow i}}} \right) = \frac{1}{2}{\sum\limits_{n = 1}^{T/2 - 1} {\left( {{{\bm B}_{1k}}\left( {{\omega _n}} \right){\cA_{in}}} \right)} ^2} .
\end{equation}
\noindent

\noindent
Then the amplification rate of the demand sequence generated from node $i$ to node $k$ is

\vspace{-6pt}
\begin{equation} \label{eq:node_to_node_BWE}
   \Phi_{{k \leftarrow i}}= \sqrt {\frac{{{\sum\limits_{n = 1}^{T/2 - 1} {\left( {{{\bm B}_{1k}}\left( {{\omega _n}} \right){\cA_{in}}} \right)^2} }}}{{\sum\limits_{n = 1}^{T/2 - 1} {\cA_{in}^2} }}} .
\end{equation}








\subsection{Impact of Network Structure and Market Demand on BWE}

Using the characterization of BWE, we establish the following results about the joint impact of network structure and market demand on the BWE of supply networks.

\begin{proposition} 
\label{thm:same_stationary_demand}
If the sequences of market demand are sampled from the same stationary process, then the average layer-wise BWE is not impacted by the layer width.
\end{proposition}


\begin{proof} [Proof] 
Suppose the layer-${l}$ output of its input demand $y_{l-1,m}$ is $y_{lm}$ $(m=1,\ldots, M)$. Because market demand sequences are generated from the same stationary process, 
the market demand sequences are independent and have the same variance. It is obvious that the subsequent output sequences after each layer are also independent. 
Due to homogeneity of nodes in the same layer, the amplification rate of each node in the layer $l$ is the same, denoted by $\phi_{l*}$, therefore the layer-${l}$ BWE given $M$ market demand sequences can be expressed as

\vspace{-6pt}
\begin{equation} \label{eq:same_stationary_demand_final}
  \E \left[ \left({\Phi _{{l}}^{(M)}} \right)^2 \right] = \E \left[ {\frac{{{ { \var }}\left( {\sum\limits_{m = 1}^M {{y_{l m}}} } \right)}}{{{ { \var }}\left( {\sum\limits_{m = 1}^M {{y_{l-1,m}}} } \right)}}} \right] = \E \left[ {\frac{{\sum\limits_{m = 1}^M {{ { \var }}\left( {{y_{lm}}} \right)} }}{{\sum\limits_{m = 1}^M {{ { \var }}\left( {{y_{l-1,m}}} \right)} }}} \right] = \E\left[ {\frac{{M \cdot \phi _{l*}^2 \cdot {\mathop{\rm var}} \left( {{y_{l-1,m}}} \right)}}{{M \cdot {\mathop{\rm var}} \left( {{y_{l-1,m}}} \right)}}} \right] = {\phi _{l*}^2}.
\end{equation}

\noindent Because layer width $M$ is not involved in the last expression in Eq.~\eqref{eq:same_stationary_demand_final}, the average layer-wise BWE is not impacted by the layer width under the given conditions.
\end{proof}

\begin{rmk}
Proposition~\ref{thm:same_stationary_demand} implies that if the sequences of market demand are generated from the same stationary process, then the layer-wise BWE of any supply networks can be reduced to that of a linear supply network subject to the same market demand, i.e., layer-wise BWE is not impacted by the network structure.
\end{rmk}

\begin{proposition} 
\label{thm:diff_stationary_demand} 
If the sequences of market demand follow stationary AR(1) processes with parameters drawn IID from a common hyperparameter distribution, then the average layer-wise BWE is monotonically decreasing as the layer width increases.
\end{proposition}

\begin{proof}[Proof] 
Suppose the stationary AR(1) process in market demand node $i$ has parameter $\varphi_i$ ($|\varphi_i|<1$)
and the associated Gaussian white noise has variance $\sigma^2$.
Because the sequences of market demand are generated from different stationary AR(1) processes, the market demand sequences are independent and their variances are different, therefore the subsequent demand sequences into each layer are also independent and have different variances. 

From Eq. \eqref{eq:link_dynamics_chain_simple}, we have

\vspace{-6pt}
\begin{equation}
y_{1i}(t)=\left(\frac{L}{P}+1\right)y_{0i}(t-1)-\frac{L}{P}y_{0i}(t-P-1)
\end{equation} 

\noindent
and therefore, let

\vspace{-6pt}
\begin{equation}
\eta_i =\frac{\mathrm{var}(y_{1i})}{\mathrm{var}(y_{0i})}=\left(\frac{L}{P}+1\right)^2+\frac{L^2}{P^2}-2\frac{L}{P}\left(\frac{L}{P}+1\right)\varphi_i^{P}.
\end{equation}

\noindent
One can easily verify that $\eta_i > 1$ and $\eta_i$ is negatively correlated $1/(1-\varphi_i^2)$. Let $\alpha_{i}=1/(1-\varphi_{i}^{2}) \ge 1$, then the mean square of layer-${l}$ BWE given $M$ market demand sequences is expressed as

\vspace{-6pt}
\begin{equation}
    \E \left[ \left({\Phi _{1}^{(M)}} \right)^2 \right] = \E\left[ {\frac{{{\rm{var}}\left( {{y_{11}} + \ldots + {y_{1M}}} \right)}}{{{\rm{var}}\left( {{y_{01}} + \ldots + {y_{0M}}} \right)}}} \right] = \E\left[ {\frac{{{\rm{var}}\left( {{y_{11}}} \right) + \ldots + {\rm{var}}\left( {{y_{1M}}} \right)}}{{{\rm{var}}\left( {{y_{01}}} \right) + \ldots + {\rm{var}}\left( {{y_{0M}}} \right)}}} \right] = 
    \E\left[ \frac{\sum\limits_{i = 1}^M \alpha_i\eta_i} {\sum\limits_{i = 1}^M \alpha_i} \right].
\end{equation}
Next we show that $\E\left[ \sum\limits_{i = 1}^M \alpha_i\eta_i /\sum\limits_{i = 1}^M \alpha_i \right]$ is monotonically decreasing as $M$ increases. Let $f(\alpha)=\alpha\eta(\alpha)$, where 

\vspace{-6pt}
\begin{equation}
\eta (\alpha) =\left(\frac{L}{P}+1\right)^2+\frac{L^2}{P^2}-2\frac{L}{P}\left(\frac{L}{P}+1\right)(1-\alpha^{-1})^{P/2}.    
\end{equation}
\noindent
When $P=1$, $f$ is convex. When $P=2$, $f$ is linear. When $P \ge 3$, $f$ is concave. Under all the three cases, the corresponding condition in Lemma~\ref{lemma: sufficient-condition} (in the next page) is satisfied. Therefore, the result is direct consequence of Lemma~\ref{lemma:ratio}.
\end{proof}

\begin{lemma}\label{lemma:ratio}
Suppose $\{(A_i, B_i)\}_{i=1}^\infty$ is an IID sequence of correlated pairs of positive random variables. If 
$$\frac{\mathrm{cov}(A, B)}{\mathrm{var}(B)}\leq \frac{\E[A]}{\E[B]},$$ we have $Z_M = \sum_{i=1}^M A_i/\sum_{i=1}^M B_i,\ M=1,2,\dots$ is monotonically decreasing for $M \geq M_0 \ge 1$ with $M_0$ being a constant.
\end{lemma}
\begin{proof}
Consider the function of two random variables $f(U, V) = U/V$. Its Taylor expansion around the point $(\mu_U, \mu_V)$ corresponds to
\vspace{-6pt}
\begin{subequations}
\begin{align}
   f(U, V) &= \frac{\mu_U}{\mu_V}+\frac{1}{k!}\sum_{k=1}^\infty\sum_{r+s=k} \frac{\partial^k f}{\partial^r U\,\partial^s V}(U-\mu_U)^r(V-\mu_V)^s\\
   &=\frac{\mu_U}{\mu_V}+\sum_{k=1}^\infty (-1)^k\mu_V^{-k}\left(\frac{\mu_U}{\mu_V}(V-\mu_V)^k-(V-\mu_V)^{k-1}(U-\mu_U)\right), 
\end{align}
\end{subequations}

\noindent
where $\mu_V=\E[V]$ and $\mu_U=\E[U]$. The expectation of $f$ is therefore

\vspace{-6pt}
\begin{equation}
\E[f(U, V)] = \frac{\mu_U}{\mu_V}+\sum_{k=2}^\infty (-1)^k\mu_V^{-k}\left(\frac{\mu_U}{\mu_V}\E\left[(V-\mu_V)^k \right]-\E\left[(V-\mu_V)^{k-1}(U-\mu_U)\right]\right).
\end{equation}

\noindent
By plugging in $U=M^{-1}\sum_{i=1}^M A_i$ and $V=M^{-1}\sum_{i=1}^M B_i$, we have

\vspace{-6pt}
\begin{equation}
\E[Z_M] = \frac{\E[A]}{\E[B]}+\sum_{k=2}^\infty (-1)^k M^{1-k} \E[B]^{-k} \left(\frac{\E[A]}{\E[B]}\E\left[(B-\E[B])^k\right]-\E\left[(B-\E[B])^{k-1}(A-\E[A])\right]\right).
\end{equation}

\noindent
Since the $k$-th term ($k\ge 2$) is proportional to $M^{1-k}$, which is strictly decreasing as $M$ increases, we choose a sufficiently large $M_0$ such that $k=2$ dominates for $M \geq M_0$. Now consider the $k=2$ term, since $\mathrm{cov}(A, B)/\mathrm{var}(B) \le \E[A]/\E[B]$, we have

\vspace{-6pt}
\begin{equation}
\E[B]^{-2}\left(\frac{\E[A]}{\E[B]}\E\left[(B-\E[B])^2\right]-\E\left[(B-\E[B])(A-\E[A])\right]\right)\geq 0.
\end{equation}


\noindent
Therefore, $\E[Z_M]$ is monotonically decreasing as $M$ increases for $M \geq M_0$.
\end{proof}


In order to easily verify the condition in Lemma~\ref{lemma:ratio}, we provide the following sufficient criterion. 
\begin{lemma}\label{lemma: sufficient-condition}
Suppose $X$ is a non-singleton random variable supported on interval $\left[e_0,\infty)\right.$ and $f$ is a differentiable function on $[e_0, \infty)$. Under either of the following two conditions: (1) $f$ is concave and $f(e_0)\geq e_0f'(e_0)$; (2) $f$ is convex and $\lim_{h\rightarrow\infty}\ h^{-1}f(h)-f'(h)\geq 0$, we have
$$\frac{\mathrm{cov}(X, f(X))}{\mathrm{var}(X)}\leq \frac{\E[f(X)]}{\E[X]}.$$
\end{lemma}

\begin{proof}
Consider the continuous function 

\vspace{-6pt}
\begin{equation}
g(h)=f(h)-\E[f(X)] - \frac{\mathrm{cov}(X, f(X))}{\mathrm{var}(X)}(h-\E[X]).
\end{equation}

\noindent
It is obvious that $\E[g(X)]=0$. We first show that there exists $h_1\in(e_0, e_1)$ such that $g(h_1)=0$. Assume the opposite and without loss of generality, assume $g(h) > 0$ for all $h\in(e_0, e_1)$. Then $\E[g(X)] > 0$, which contradicts $\E[g(X)]=0$. Therefore, $h_1\in(e_0, e_1)$ exists. 

Furthermore, suppose $h_1$ is the only zero of $g(h)$. By the continuity of $g$ and without loss of generality, assume $g(h) < 0$ for $h<h_1$ and $g(h) > 0$ for $h > h_1$, then we have $g(h)(h-h_1)>0$ for all $h$. After careful calculation one would find that $\E[g(X)(X-h_1)]=0$, which leads to contradiction again. Therefore, there exists another point $h_2\neq h_1$ such that $g(h_2)=0$. 

Without loss of generality, we assume $h_1<h_2$.
As $f(h)-g(h)$ is a linear function passing $(h_1, f(h_1))$ and $(h_2, f(h_2)$, we have

\vspace{-6pt}
\begin{equation}
\E[f(X)] + \frac{\mathrm{cov}(X, f(X))}{\mathrm{var}(X)}(h-\E[X])=f(h_1) + \frac{f(h_2)-f(h_1)}{h_2-h_1}(h-h_1).
\end{equation} 

\noindent
Let $h=0$, then we have

\vspace{-6pt}
\begin{equation}
\E[f(X)] - \frac{\mathrm{cov}(X, f(X))}{\mathrm{var}(X)}\E[X]=f(h_1) - \frac{f(h_2)-f(h_1)}{h_2-h_1}h_1.
\end{equation}

\noindent
When $f$ is concave, $f(h_2)-f(h_1)/(h_2-h_1)\leq f'(h_1).$ Therefore

\vspace{-6pt}
\begin{equation}
\E[f(X)] - \frac{\mathrm{cov}(X, f(X))}{\mathrm{var}(X)}\E[X]\geq f(h_1)-h_1 f'(h_1)\geq f(e_0)-e_0f'(e_0)\geq 0.
\end{equation}

\noindent
The result is now straightforward. The proof for the case where $f$ is convex is similar and is thus omitted here.
\end{proof}

\begin{rmk}
We note that Lemma~\ref{lemma: sufficient-condition} provides sufficient and necessary condition for Lemma~\ref{lemma:ratio} to hold for arbitrary distribution of $(A, B)$. The concluded trend in Lemma~\ref{lemma:ratio} can be observed under specific distributions of $(A, B)$, even when Lemma~\ref{lemma: sufficient-condition} is violated.   
The conditions in Lemma~\ref{lemma: sufficient-condition} are considered under the worst-case of all possible distributions of $X$. Specifically, condition (1) is taken when the distribution of $X$ is focused around the point $a$ while condition (2) is considered when the probability densities of $X$ are allocated to large numbers.
\end{rmk}

\begin{proposition} 
\label{thm:non_stationary_demand}
Suppose the sequences of market demand are non-stationary processes such that
$$y_{0i}=c+at+\tilde y_{0i},$$
where $\tilde y_{0i}$ follows a stationary AR(1) process with its coefficient $\varphi$ drawn IID from a common distribution.
Then the average layer-wise BWE with the moving average window size $P\leq 2$ decreases as the layer width increases.
\end{proposition}

\begin{proof} [Proof] 
According to Eq. \eqref{eq:link_dynamics_chain_simple}, we have

\vspace{-6pt}
\begin{subequations}
\begin{align}
c+at + \tilde y_{1i}(t)&=\left(\frac{L}{P}+1\right)\left(c+a(t-1)+\tilde y_{0i}(t-1)\right)-\frac{L}{P}\left(c+a(t-P-1)+\tilde y_{0i}(t-P-1)\right)\\
&=\left(\frac{L}{P}+1\right)\tilde y_{0i}(t-1)-\frac{L}{P}\tilde y_{0i}(t-P-1)+a(L+t-1)+c,
\end{align}
\end{subequations}

\noindent
which gives
\vspace{-6pt}
\begin{equation}
\tilde y_{1i}(t)=\left(\frac{L}{P}+1\right)\tilde y_{0i}(t-1)-\frac{L}{P}\tilde y_{0i}(t-P-1)+a(L-1).
\end{equation}

\noindent
Therefore,

\vspace{-6pt}
\begin{equation}
\frac{\mathrm{var}(\tilde y_{1i})}{\mathrm{var}(\tilde y_{0i})} = \left(\frac{L}{P}+1\right)^2+\frac{L^2}{P^2}-2\frac{L}{P}\left(\frac{L}{P}+1\right)\varphi_i^{P}=:\tilde\eta_i.
\end{equation}

\noindent
Notice that $y_{1i}=c+at+\tilde y_{1i}$ and $y_{0i}=c + at + \tilde y_{0i}$, then we have

\vspace{-6pt}
\begin{equation} \label{eq:ratio_with_trend}
\eta_i:=\frac{\mathrm{var}(y_{1i})}{\mathrm{var}(y_{0i})}=\frac{\alpha_i\tilde\eta_i + \tau}{\alpha_i + \tau},
\end{equation}
\noindent
where $\alpha_i= 1/(1-\varphi_i^2)$ is the variance of AR(1) process and $\tau=a^2(T^2-1)/12$ is the variance of the linear trend term ($at$). 
One can then follow the same argument in the proof of Proposition~\ref{thm:diff_stationary_demand}. Condition in Lemma~\ref{lemma: sufficient-condition} is satisfied only when $P \le 2$.
\end{proof}

\begin{rmk}
  The condition $P\leq 2$ in Proposition~\ref{thm:non_stationary_demand} ensures the decreasing trend of layer-wise BWE over the increase in layer width holds for an arbitrary distribution for $\varphi$ when $P\leq 2$. When $P\geq 3$, which is more often the case in practice, the trend of layer-wise BWE depends on the specific distribution for AR coefficients. To determine the trend for given distributions, one should verify the condition in Lemma~\ref{lemma:ratio}. Theoretically, for $P\geq 3$, the layer-wise BWE can be increasing as the layer width increases if the worst-case scenario in condition (1) in Lemma~\ref{lemma: sufficient-condition} is violated. 
  But for general distributions of $\varphi$ that spread out the support $(-1, 1)$, one can usually observe a decreasing trend of layer-wise BWE as the layer width increases.
\end{rmk}


\section{Numerical Experiments}

\subsection{Experimental Setup}


We use numerical experiments to validate the proposed method for characterizing BWE and the subsequent proposals. In the experiments, order sequences from suppliers are then generated from different market demand sequences according to Eq. \eqref{eq:dynamics_net}. For the impact of stochastic lead times on BWE, readers are referred to \citep{kim2006quantifying}.
The number of replications is 50 while the number of periods is 1000 with the first 400 as warm-ups. These numbers are chosen to ensure sufficiently small variations in the results. The weights of outgoing links of each node are assumed to be equal in order to focus the analysis on network structure.

\subsubsection{Market Demand Patterns}
We consider both stationary and non-stationary market demand patterns. In validating the methodology for calculating layer-wise BWE, demand patterns are simulated using a flexible model for time series data given by

\vspace{-6pt}
\begin{equation} \label{eq:market_demand}
    {y_t} = c + at 
    + \sum\limits_{n = 1}^h {{\gamma _n}} \sin \left( {2\pi  \cdot {v_n}t} \right) + \cN \left(0, \sigma^2 \right) ,\; t=1.\ldots, T,
    \vspace{-6pt}
\end{equation}

\noindent
where $c$ is the base level. $at$ is the trend component (exogenous factor). 
$\sum\limits_{n = 1}^h {{\gamma _n}} \sin \left( {2\pi  \cdot {v_n}t} \right)$ is the seasonality component where $\gamma_n$ and $v_n$ are the respective amplitude and frequency~\citep{pollock1993lectures} that allow us to model long- and short-term temporal patterns. A similar model is used in \citep{jakvsivc2008effect}.
We generate 4 specific market demand patterns shown in column 3 of Table \ref{tab:diff_structure_demand}. Demand pattern 1 is stationary. Demand patterns 2 to 4 are non-stationary, in which demand pattern 2 emphasizes the seasonal fluctuations, demand pattern 3 emphasizes the increasing trend and seasonal fluctuations, and demand pattern 4 emphasizes increasing demand even more. For market demand pattern generated with an increasing trend component, the layer-wise BWE becomes

\vspace{-6pt}
\begin{equation} \label{eq:layerwise-bwe-nonstation-general}
    {\Phi_l} = \sqrt {\frac{{\frac{1}{2}\sum\limits_{i \in \cL_l} {\sum\limits_{n = 1}^{T/2 - 1} {\left( {\phi _n^{2l}\cA_n^2} \right)}  + \tau    {{\left| {\cL_1} \right|}^2}} }}{{\frac{1}{2} \sum\limits_{i \in \cL_l} {\sum\limits_{n = 1}^{T/2 - 1} {\left( {\phi _n^{2\left( {l - 1} \right)}{\cA}_n^2} \right)}  + \tau    {{\left| {\cL_1} \right|}^2}} }}} ,\; \forall\, {l} \in \cL,
\end{equation}


\noindent
where $\left| {\cL_1} \right|$ is the number of market demand sequences into layer one. $\tau = a^2\left(T^2-1\right) / 12$ is the variance of the trend component $at$. Note that the exogenous component $at$ will not lead to variance amplification. For example, trend component represented by $a t$ in linear supply networks with dynamics as in Eq. \eqref{eq:link_dynamics_chain_simple} lead to demand sequence with the same variance after going through a node. Formally, ${y_{12}}\left( t \right) = {l_1} \cdot \frac{{a\left( {t - 1} \right) - a\left( {t - {r_1} - 1} \right)}}{{{r_1}}} + a\left( {t - 1} \right) = at + a\left( {{l_1} - 1} \right)$, 
then $ \var\left({y_{12}}\right) = \var\left( \{at\}^{T}_{t=1} \right) + 0= \var\left({y_{01}}\right)$, therefore the variance of demand does not amplify.

In validating the propositions, the corresponding stationary or non-stationary demand processes will be used. The details about each demand process and other parameters will be provided along with the associated results.

\begin{figure}[!htb]
    \centering
    \includegraphics[width=0.85\textwidth]{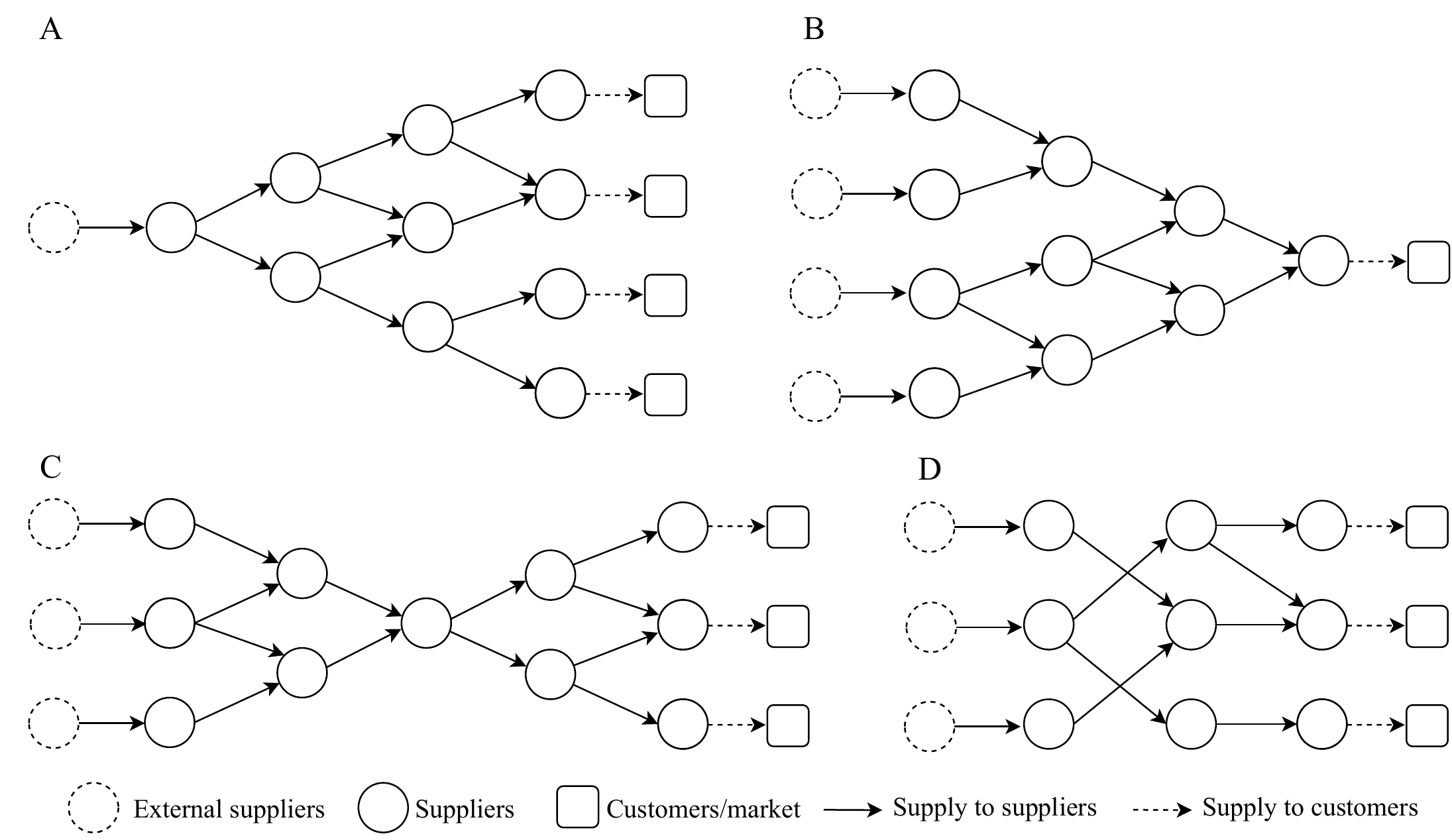}
    \caption{Schematic of different types of supply network structures. A. Divergent (Div). B. Convergent (Conv). C. Divergent-to-Convergent (Div2Conv). D. Parallel (Paral).}
    \label{fig:structure_types}
\end{figure}

\subsubsection{Network Structures}

To validate the impact of network structure on layer-wise BWE, we use several disparate types of network structures, including divergent (Div) network wherein the layer width increases linearly from upstream to downstream, convergent (Conv) network wherein the layer width decreases linearly from upstream to downstream, divergent-then-convergent (Div2Conv) network wherein the layer width decreases linearly first then increases linearly from upstream to downstream, and parallel (Paral) network wherein all layers have the same width (Fig. \ref{fig:structure_types}). All the networks have unidirectional flow between suppliers at different layers and there exists at least one link from and to any suppliers in intermediate layers, one link from suppliers in the final layer, and one link to suppliers in the first layer. Customers (market demand) are positioned in layer zero (virtual layer). We use a parameter, $\rho$ to model networks with various inter-layer link densities. $\rho$ is defined as the probability of a link between a node $i$ and all nodes in the immediate downstream layer but the first one already connected to node $i$. This ensures that: when $\rho=1$, an upstream node is connected to all nodes in its immediate downstream layer; when $\rho=0$, an upstream node is connected to only one node in the immediate downstream layer.

\subsection{Numerical Results} \label{sec:result}   
\begin{figure}[!htb]
    \centering
    \includegraphics[width=0.85\textwidth]{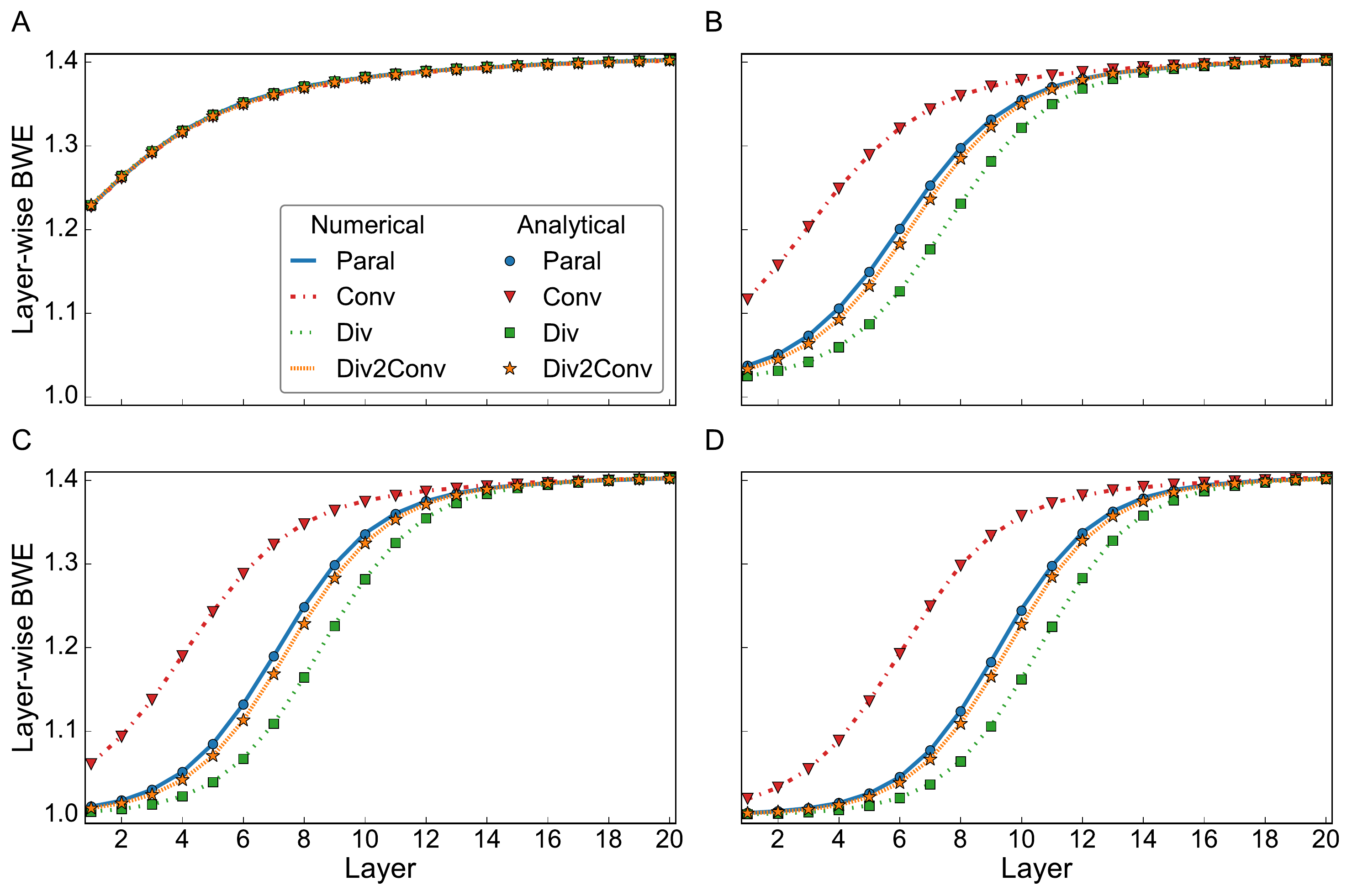}
    \caption{Layer-wise BWEs of different supply networks across different market demands using numerical simulation and analytical analysis. Figures A to D are generated using market demand patterns 1 to 4 in Table \ref{tab:diff_structure_demand}, respectively. The lead time $L_i$ is 4 and the moving average window $P_i$ is 19 following \citet{dejonckheere2004impact}.}
    \label{fig:num_vs_analy}
\end{figure}

To validate the characterization of layer-wise BWE, the simulated demand patterns and the respective RMSE of layer-wise BWE are presented in Table \ref{tab:diff_structure_demand}. The RMSE of layer-wise BWEs is calculated by $\sqrt{\frac{\sum_{l=1}^{\left| \cL\right|}\left(\Phi_{l}-\hat{\Phi}_{l}\right)^{2}}{\left| \cL\right|}}$ where $\hat{\Phi}_{l}$ is the estimated BWE of layer $l$. From Table \ref{tab:diff_structure_demand} we can observe that under different market demand patterns and network structures, the RMSE of layer-wise BWE is very small, indicating that the analytical solutions are very close to the numerical results. The comparison between layer-wise BWEs obtained by numerical experiments and analytical solutions in Fig. \ref{fig:num_vs_analy} also confirms the high accuracy of the analytical characterization of layer-wise BWE of different supply networks. From Fig. \ref{fig:num_vs_analy} A, where the market demand sequences are generated from the same stationary distribution, we can see that the layer-wise BWEs are the same across different types of network structures. This result validates Proposition \ref{thm:same_stationary_demand} since the layer widths in the same layer for different network structures are not all the same. In contrast, when the market demand sequences are generated from non-stationary process (B, C, and D), the layer-wise BWEs are no longer the same for different network structures.

\begin{table}[htb]
  \centering
  \caption{BWE for different network structures under different demand patterns}
    \begin{tabular}{ccccc}
    \toprule
    NO.    & \multicolumn{1}{p{8.2em}}{\centering Network structure} & \multicolumn{2}{c}{Market demand pattern} & \multicolumn{1}{p{9.5em}}{\centering RMSE of layer-wise BWE ($\times 10 ^{-7}$)} \\
    \midrule
    \multirow{10}[2]{*}{1} & \multicolumn{1}{c}{\multirow{2}[1]{*}{Paral}} & \multicolumn{2}{c}{$y_t = 100 + \cN \left(0, 20^2 \right)$ } & \multicolumn{1}{c}{\multirow{2}[1]{*}{2.06}} \\
          &       & \multicolumn{2}{c}{\multirow{9}[1]{*}{\raisebox{-\totalheight}{\vspace{-35pt} \includegraphics[width=0.25\textwidth]{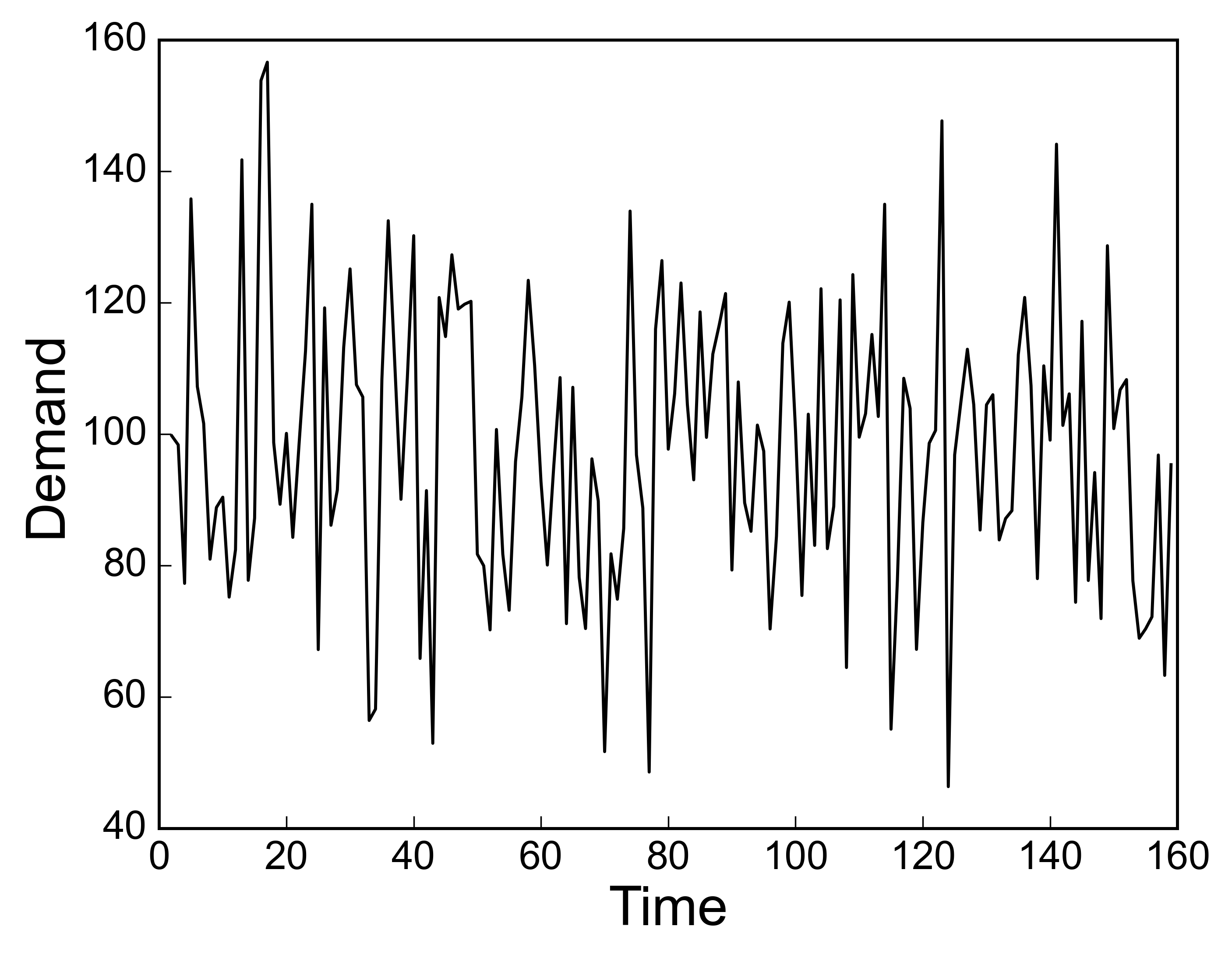}}}} &  \\
          & \multicolumn{1}{c}{\multirow{2}[0]{*}{Conv}} & \multicolumn{2}{c}{} & \multicolumn{1}{c}{\multirow{2}[0]{*}{2.91}} \\
          &       & \multicolumn{2}{c}{} &  \\
          & \multicolumn{1}{c}{\multirow{2}[0]{*}{Div}} & \multicolumn{2}{c}{} & \multicolumn{1}{c}{\multirow{2}[0]{*}{2.20}} \\
          &       & \multicolumn{2}{c}{} &  \\
          & \multicolumn{1}{c}{\multirow{2}[0]{*}{Div2Conv}} & \multicolumn{2}{c}{} & \multicolumn{1}{c}{\multirow{2}[0]{*}{1.22}} \\
          &       & \multicolumn{2}{c}{} &  \\
          &       & \multicolumn{2}{c}{} &  \\
          &       & \multicolumn{2}{c}{} &  \\
    \midrule
    \multirow{8}[1]{*}{2} & \multicolumn{1}{c}{\multirow{2}[1]{*}{Paral}} & \multicolumn{2}{c}{$y_t = 100 + 10 \sin (2 \pi \cdot 0.1 t) + 30 \sin(2 \pi \cdot 0.05 t)+\cN \left(0, 20^2 \right)$} & \multicolumn{1}{c}{\multirow{2}[1]{*}{1.14}} \\
          &       & \multicolumn{2}{c}{\multirow{9}[1]{*}{\raisebox{-\totalheight}{\vspace{-35pt} \includegraphics[width=0.25\textwidth]{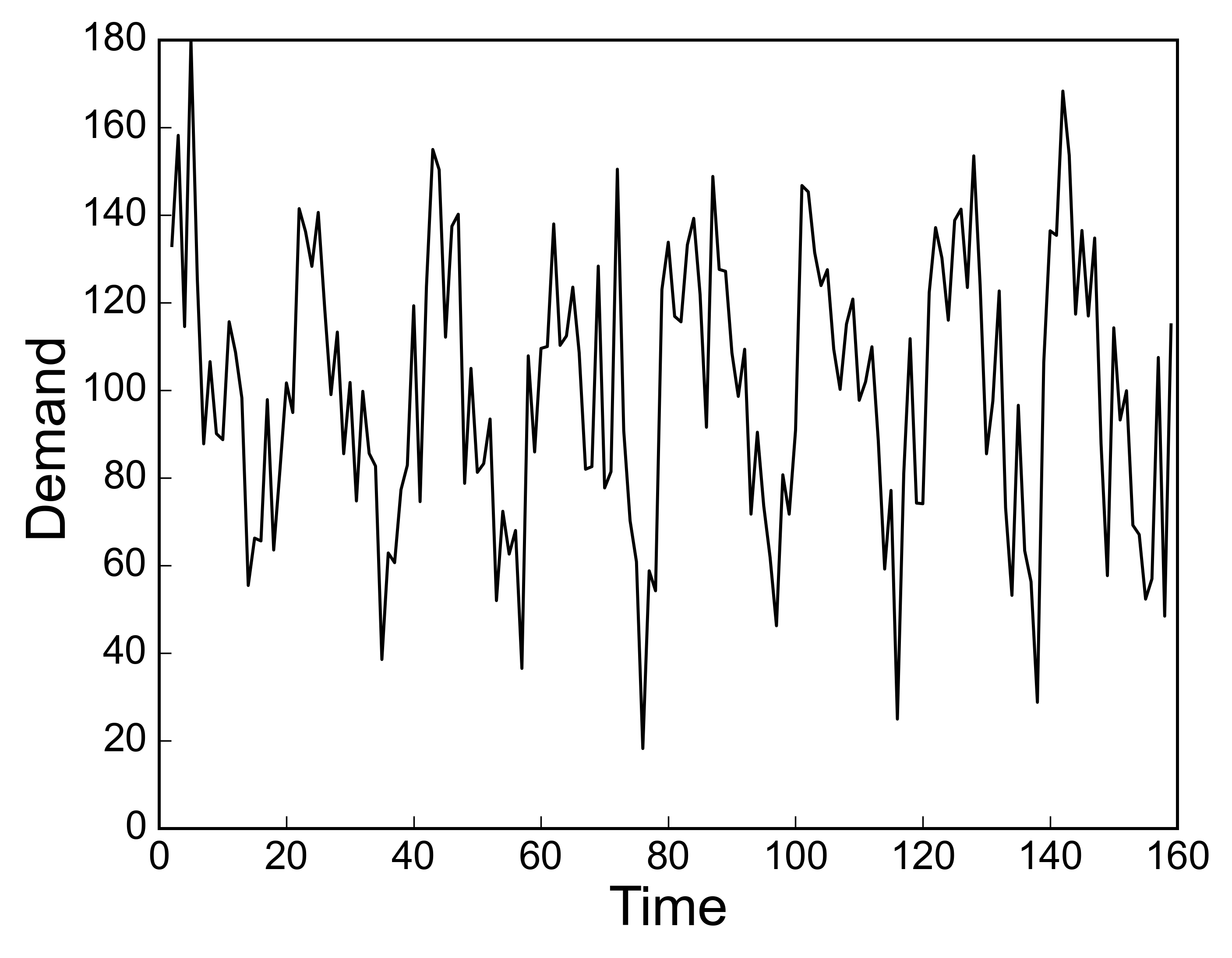}}}} &  \\
          & \multicolumn{1}{c}{\multirow{2}[0]{*}{Conv}} & \multicolumn{2}{c}{} & \multicolumn{1}{c}{\multirow{2}[0]{*}{2.12}} \\
          &       & \multicolumn{2}{c}{} &  \\
          & \multicolumn{1}{c}{\multirow{2}[0]{*}{Div}} & \multicolumn{2}{c}{} & \multicolumn{1}{c}{\multirow{2}[0]{*}{1.79}} \\
          &       & \multicolumn{2}{c}{} &  \\
          & \multicolumn{1}{c}{\multirow{2}[0]{*}{Div2Conv}} & \multicolumn{2}{c}{} & \multicolumn{1}{c}{\multirow{2}[0]{*}{2.26}} \\
          &       & \multicolumn{2}{c}{} &  \\
          &       & \multicolumn{2}{c}{} &  \\
          &       & \multicolumn{2}{c}{} &  \\
    \midrule
    \multirow{10}[2]{*}{3} & \multicolumn{1}{c}{\multirow{2}[1]{*}{Paral}} & \multicolumn{2}{c}{$y_t = 100 + 0.2 t + 10 \sin (2 \pi \cdot 4 t) + 20 \sin(2 \pi \cdot 0.5 t)+\cN \left(0, 20^2 \right)$
    } & \multicolumn{1}{c}{\multirow{2}[1]{*}{4.14}} \\
          &       & \multicolumn{2}{c}{\multirow{9}[1]{*}{\raisebox{-\totalheight}{\vspace{-35pt} \includegraphics[width=0.25\textwidth]{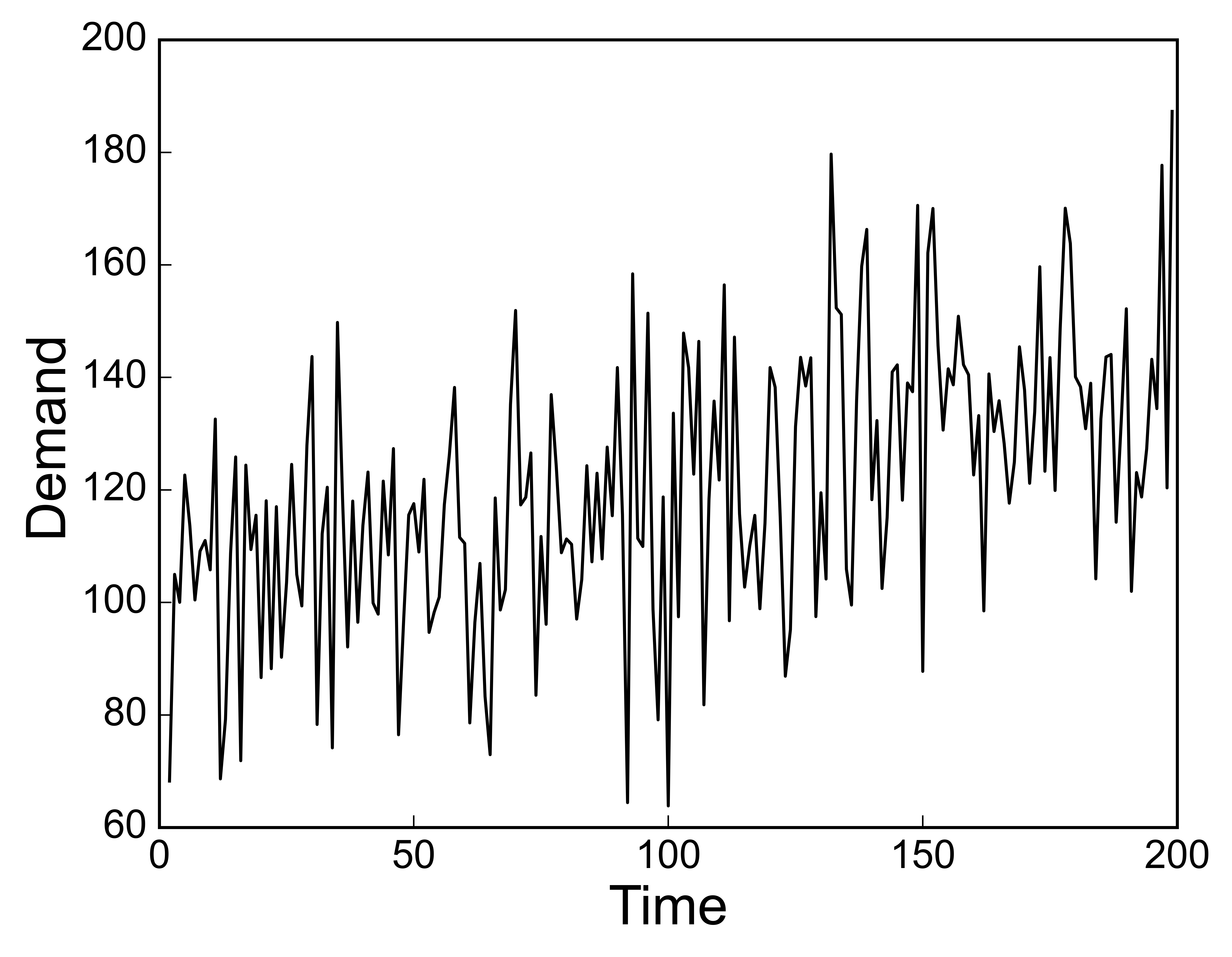}}}} &  \\
          & \multicolumn{1}{c}{\multirow{2}[0]{*}{Conv}} & \multicolumn{2}{c}{} & \multicolumn{1}{c}{\multirow{2}[0]{*}{1.11}} \\
          &       & \multicolumn{2}{c}{} &  \\
          & \multicolumn{1}{c}{\multirow{2}[0]{*}{Div}} & \multicolumn{2}{c}{} & \multicolumn{1}{c}{\multirow{2}[0]{*}{8.14}} \\
          &       & \multicolumn{2}{c}{} &  \\
          & \multicolumn{1}{c}{\multirow{2}[0]{*}{Div2Conv}} & \multicolumn{2}{c}{} & \multicolumn{1}{c}{\multirow{2}[0]{*}{7.46}} \\
          &       & \multicolumn{2}{c}{} &  \\
          &       & \multicolumn{2}{c}{} &  \\
          &       & \multicolumn{2}{c}{} &  \\
    \midrule
    \multirow{10}[2]{*}{4} & \multicolumn{1}{c}{\multirow{2}[1]{*}{Paral}} & \multicolumn{2}{c}{$y_t = 100 + 0.4 t + 10 \sin (2 \pi \cdot 4 t) + 10 \sin(2 \pi \cdot 2 t)+\cN \left(0, 20^2 \right)$} & \multicolumn{1}{c}{\multirow{2}[1]{*}{6.27}} \\
          &       & \multicolumn{2}{c}{\multirow{9}[1]{*}{\raisebox{-\totalheight}{\vspace{-35pt} \includegraphics[width=0.25\textwidth]{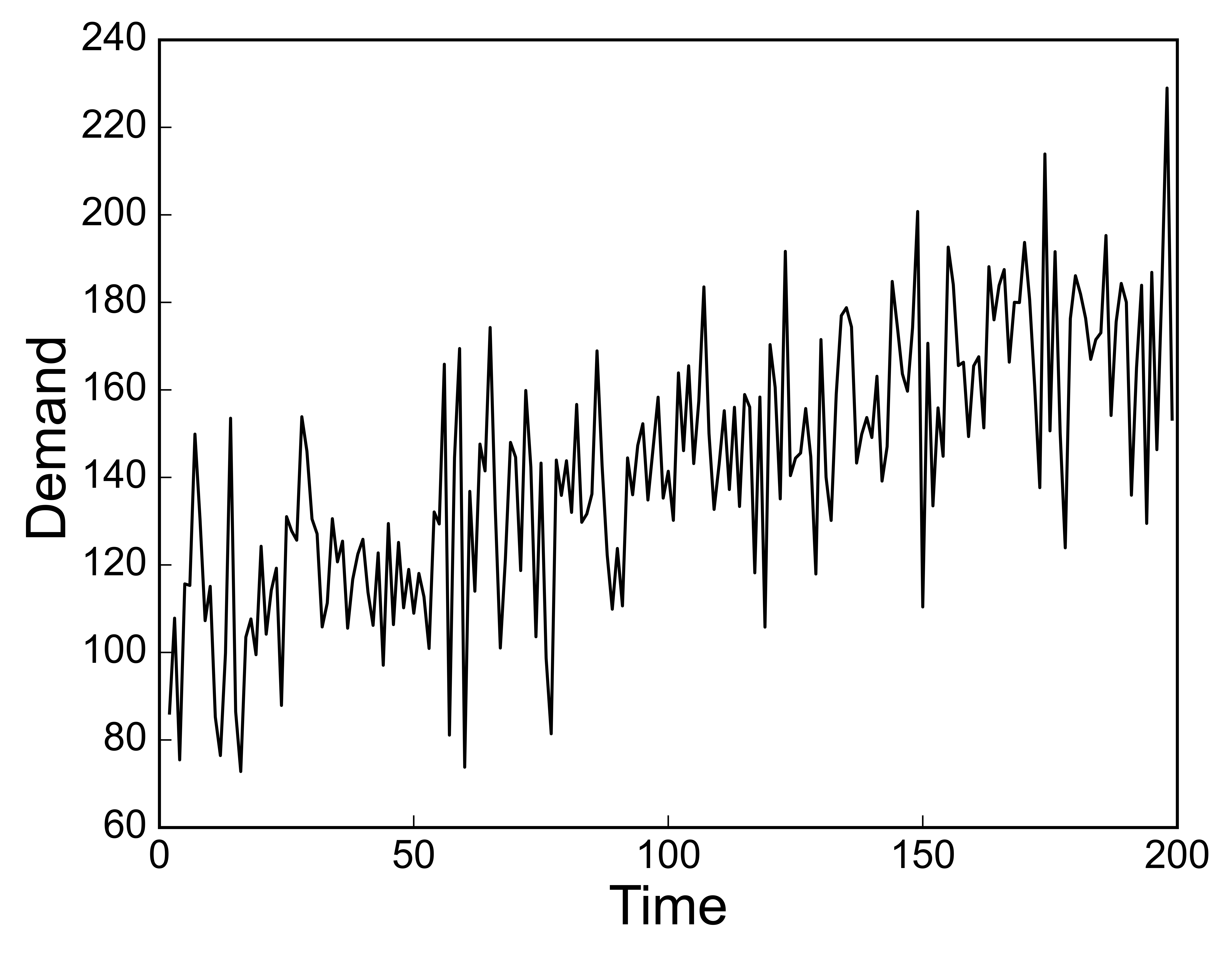}}}} &  \\
          & \multicolumn{1}{c}{\multirow{2}[0]{*}{Conv}} & \multicolumn{2}{c}{} & \multicolumn{1}{c}{\multirow{2}[0]{*}{1.84}} \\
          &       & \multicolumn{2}{c}{} &  \\
          & \multicolumn{1}{c}{\multirow{2}[0]{*}{Div}} & \multicolumn{2}{c}{} & \multicolumn{1}{c}{\multirow{2}[0]{*}{4.53}} \\
          &       & \multicolumn{2}{c}{} &  \\
          & \multicolumn{1}{c}{\multirow{2}[0]{*}{Div2Conv}} & \multicolumn{2}{c}{} & \multicolumn{1}{c}{\multirow{2}[0]{*}{1.46}} \\
          &       & \multicolumn{2}{c}{} &  \\
          &       & \multicolumn{2}{c}{} &  \\
          &       & \multicolumn{2}{c}{} &  \\
    \bottomrule
    \end{tabular}%
  \label{tab:diff_structure_demand}%
\end{table}%

\begin{figure}[!htb]
    \centering
    \includegraphics[width=0.85\textwidth]{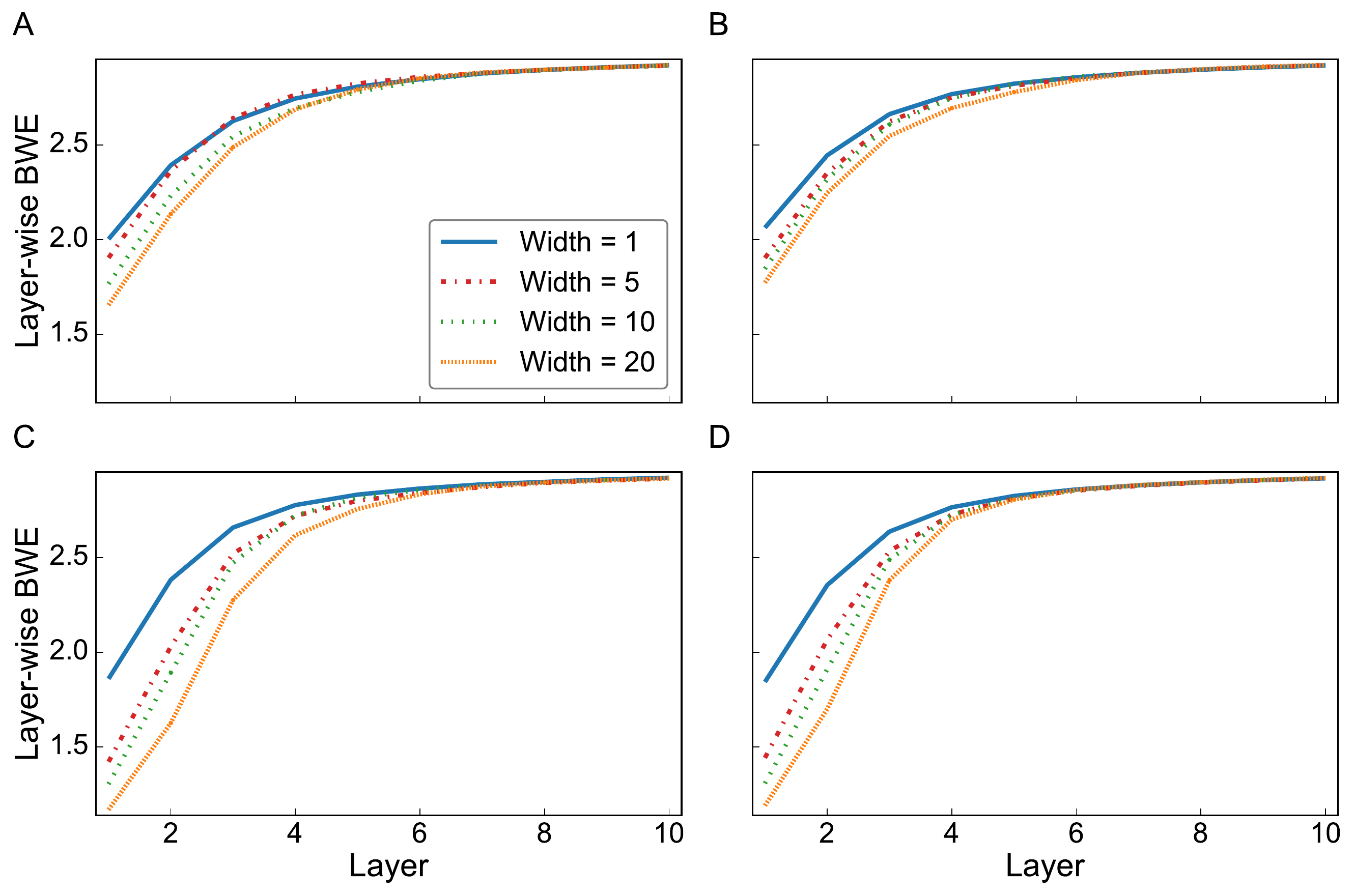}
    \caption{Layer-wise BWE of parallel supply networks across different layer widths and demand sequences. A. $a=0$ and $\varphi \sim U(-1,1)$. B. $a=0$ and $\varphi$ follows a truncated normal distribution $\bar{\cN} (0,1,-1,1)$ with the mean being 0, variance 1, lower bound -1, and upper bound 1. c. $a=0.1$ and $\varphi \sim U(-1,1)$. D. $a=0.1$ and $\varphi \sim \bar{\cN} (0,1,-1,1)$. The layer width is 8 and the link probability is 0.25. $L_i=P_i=4$.}
    \label{fig:bwe_over_width}
\end{figure}

To verify the propositions (\ref{thm:diff_stationary_demand} and \ref{thm:non_stationary_demand}) about the impact of market demand, we show the layer-wise BWE of parallel supply networks with different layer widths in Fig. \ref{fig:bwe_over_width}. We can observe that when subject to stationary AR(1) demand sequences with the parameter $\varphi$ drawn IID from a uniform distribution (A) or a truncated normal distribution (B) that spread out the interval (-1, 1), supply networks with greater layer widths have lower average layer-wise BWE. A very similar relation between layer-wise BWE and layer width can be observed when the market demand is generated from non-stationary processes composed of AR(1) process and a linear trend.

\begin{figure}[htbp]
    \centering
    \includegraphics[width=0.88\textwidth]{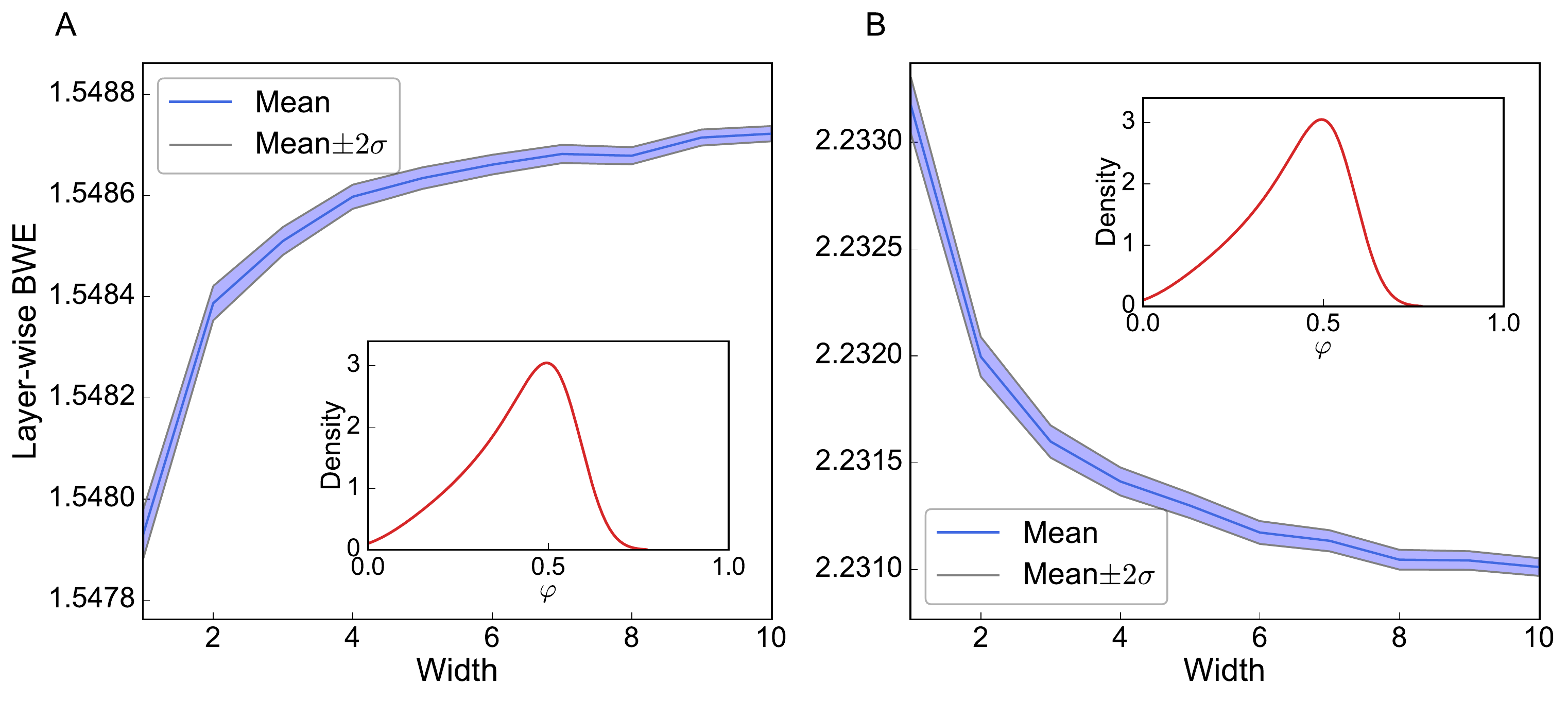}
    \caption{Trends of BWE for the first layer under different values of $P$: A. $P=2$; B. $P=4$. For both A and B, $L=2$ and the variance of linear trend is 2, i.e., $\tau=2$ in Eq. \eqref{eq:ratio_with_trend}.}
    \label{fig:bwe_width_increase_eg}
\end{figure}

To verify that layer-wise BWE can be increasing or decreasing depending on the value of $P$, we present such as example in Fig. \ref{fig:bwe_width_increase_eg}. We can observe an increasing trend of layer-wise BWE as the width increases when $P\ge3$. For the sample distribution of $\varphi$, the trend of BWE is decreasing as layer width increases if $P=2$. Note that in this specific example, the increase or decrease in layer-wise BWE due to the increase in width is not very large, but the trends are opposing.

\section{Conclusions} \label{sec:conc}

In this paper, we consider general supply networks of different network structures. We have characterized the layer-wise BWE of general supply networks using the control-theoretic approach when the layer position of a node (supplier) is unique. Additionally, we employ the absorbing Markov Chain to derive the analytical characterization of the node-wise BWE for generic supply networks, in which nodes are positioned in multiple layers due to intra-layer links among nodes or links between nodes that are not positioned in consecutive layers. We then investigate the impact of the structure of supply networks on their BWE under multiple market demand patterns. The key conclusions of our work are as follows: (i) if the market demand is generated from the same stationary process, the structure of supply networks does not affect the layer-wise bullwhip effect of supply networks; (ii) if the market demand is generated from different stationary or non-stationary market processes, wider supply networks lead to a lower level of layer-wise bullwhip effect.

This work only considers the supply network of a single end product. Future work can investigate the fluctuation propagation (particularly cascading failures) in supply networks for multiple products from the perspective of multilayer complex networks due to the interconnections among the flow of materials, information, and money. Another promising extension is to characterize the BWE of supply networks after incorporating inter–supplier interplay given only partial visibility of supply chains, such as competition, cooperation, and coopetition among suppliers \citep{bouncken2015coopetition}.


\section*{Acknowledgement}
This research was partially supported by the US National Science Foundation under grant no. 2047488. 

\section*{Data and Code Availability Statement}

Data sharing is not applicable to this paper since no new data were created. The code used for the numerical experiments in this paper will be shared publicly upon publication.

\appendix \label{sec:app}

\printcredits

\bibliographystyle{cas-model2-names}

\bibliography{main-cas-refs}

\end{document}